\documentclass[
aip,
amsmath,amssymb,
reprint,
]{revtex4-1}

\usepackage{graphicx}
\usepackage{dcolumn}
\usepackage{bm}
\usepackage[utf8]{inputenc}
\usepackage[T1]{fontenc}
\usepackage{etoolbox}
\usepackage{subfig}        
\usepackage{nicefrac}   
\usepackage{comment}
\usepackage{booktabs}
\usepackage{xcolor}         
\usepackage{lineno}
\usepackage{hyperref}
\usepackage{enumitem}
\usepackage{cleveref} 
 \usepackage{multirow}
\usepackage{amssymb} 
\usepackage{microtype}
\usepackage{algorithm}
\makeatletter
\edef\ftype@algorithm{\the\numexpr\c@float@type/2\relax}
\makeatother
\usepackage{algorithmic}
\usepackage{tikz}
\usetikzlibrary{arrows.meta,positioning,fit,calc,shapes.multipart,backgrounds}

\DisableLigatures[-]{}

\DeclareMathOperator{\softmax}{softmax}
\DeclareMathOperator{\GELU}{GELU}

\DeclareMathOperator{\LayerNorm}{LayerNorm}
\DeclareMathOperator{\FFN}{FFN}

\begin{document}

\preprint{AIP/123-QED}

\title[]{Physics Attention Transformer Surrogate for Rapid Vertical Instability Growth Rate Prediction: Alcator C-Mod to SPARC}
\author{A. Kumar}
\email{arunavk@mit.edu}

\author{C. Clauser}
\author{T. Golfinopoulos}
\author{C. Rea} 
\affiliation{Plasma Science \& Fusion Center, Massachusetts Institute of Technology, Cambridge, MA, USA }
\author{F.Capersene}
\affiliation{École Polytechnique Fédérale de Lausanne, Swiss Plasma Center, CH-1015 Lausanne, Switzerland}
\author{D. Boyer}
\affiliation{Commonwealth Fusion Systems, Cambridge, MA, USA}
\author{SPARC Team}
\affiliation{Commonwealth Fusion Systems, Cambridge, MA, USA}
\author{Alcator C-Mod Team}
 \affiliation{Plasma Science \& Fusion Center, Massachusetts Institute of Technology, Cambridge, MA, USA
}

\date{\today}

\begin{abstract}

In this work, we investigate rapid prediction of the dominant $n{=}0$ vertical instability growth rate in C-Mod and SPARC equilibria, where nonrigid free boundary response models are too slow for control cycle use. Using a Physics Attention Transformer trained on MEQ-FGE-L labels, we predict both the scalar growth rate and the associated two dimensional perturbed toroidal current density. We find mean absolute errors of 5.4~s$^{-1}$ on held out C-Mod equilibria and 12.7~s$^{-1}$ on synthetic SPARC cases, with spatial eigenfunction errors near 5\%. We also compared PAT with operator based ML models : FNO2D and DeepONet, where we found PAT predicts a much lower normalised growth rate error and improved spatial reconstruction.
These results indicate that PAT can reproduce MEQ-FGE-L outputs at control relevant latency and could support future studies of growth rate headroom monitoring and proximity aware shape control.

% Vertical stability observers must estimate instability growth rates on control cycle timescales, while nonrigid free boundary solvers require minutes per equilibrium. We present a Physics Attention Transformer (PAT) surrogate for the dominant $n{=}0$ vertical growth rate and the associated two dimensional perturbed toroidal current density. PAT compresses a high resolution equilibrium mesh into learned low dimensional tokens, appends electromagnetic coupling embeddings, applies attention in this reduced token space, and decodes both scalar and spatial outputs. The model is trained against MEQ-FGE-L calculations from Alcator C-Mod experimental equilibria and synthetic SPARC scenarios. On held out data, PAT reproduces solver growth rates with MAE $=5.4$~s$^{-1}$ on C-Mod and $12.7$~s$^{-1}$ on SPARC; spatial eigenfunction errors remain near the 5\% level. Single core inference is 35~ms, compared with ${\sim}180$~s for MEQ-FGE-L. These numbers measure surrogate to solver fidelity, not validation against nonlinear plasma dynamics. The main implication is therefore operational but bounded: PAT can serve as a fast advisory layer or triage tool, provided that uncertainty gating and independent physics based safeguards remain in the loop.

\end{abstract}
%\linenumbers
\maketitle
%\tableofcontents

\section{Introduction}

Vertical instability poses one of the practical limits of highly elongated tokamak operations. When an axisymmetric $n{=}0$ vertical mode becomes unstable, the plasma column moves toward surrounding conductors on a timescale set by passive wall shielding and feedback response\cite{Grad1967,Okabayashi_1974,lazarus_control_1990,leuer_passive_1989}. For plasma control, the relevant question is not only whether the plasma column has moved. A robust controller also needs to know how rapidly an unstable $n=0$ axisymmetric mode is growing, because that growth rate separates routine feedback action from states that require enhanced monitoring, controller reconfiguration, or disruption mitigation preparation.

Real time access to that estimate remains difficult. Reduced rigid plasma models, including RZIP style formulations, are attractive because they are fast, interpretable, and closely tied to control oriented quantities such as wall time, passive stability, and coil authority\cite{humphreys_axisymmetric_1993,Humphreys_2009,Olofsson_2022}. Their speed comes from a restrictive physical assumption: the plasma current distribution moves as a nearly fixed object. Experiments and modeling studies have shown that nonrigid current redistribution and shape response can change vertical growth rate predictions substantially, especially in shaped plasmas operated near stability limits\cite{Albanese_1989,Ward1992EffectsOP,Hofmann_1997,welander_nonrigid_2005}. Nonrigid free boundary solvers address this physics by coupling the plasma response to the external circuit and passive conductors, but that fidelity comes at a cost incompatible with control cycle evaluation.

MEQ-FGE-L code \cite{francesco2021,Heiss2025FGE} framework  belongs to this nonrigid linear response class. It computes the instantaneous plasma response around a reconstructed equilibrium, folds that response into conductor dynamics, and obtains the dominant vertical growth rate from the resulting linearized eigenproblem\cite{francesco2021,Marchioni2024}. This makes MEQ-FGE-L a useful high fidelity labeling tool for surrogate development. It is not, however, a drop in real time observer: a calculation that takes minutes per equilibrium cannot sit directly inside a control loop whose useful decision timescale is tens of milliseconds or less\cite{Ferrara2008PlasmaIA,Creely_2020,creely23,wai2025}.

The need is especially sharp for cross device planning. Alcator C-Mod provides experimental equilibria across Ohmic, L-mode, I-mode, and H-mode operation, with reconstructed magnetic states and disruption relevant trajectories\cite{Marmar_2015,Montes_2019}. SPARC occupies a different regime: compact high field operation, reactor relevant shaping, and vertical stability constraints tied to its vessel, coils, and control architecture\cite{Creely_2020,rodriguez-fernandez_overview_2022,creely23,Nelson_2024}. A surrogate trained only on scalar descriptors risks missing the spatial structure that distinguishes these machines, including plasma wall gaps, X point topology, passive conductor geometry, and the two dimensional current perturbation associated with the unstable eigenmode.

Machine learning surrogates can reduce latency, but the representation must respect the geometry of the problem. Disruption predictors and scalar input stability models have shown the value of data driven inference in tokamak operation\cite{morabito1997fuzzy,Montes_2019,decaf,app10196683,Gopakumar_2024}. Neural operators and attention based solvers offer a different route for field to field or field to scalar maps on complex geometries\cite{Li2020fno,Li2021geo,lu2021learning,Wu2024,Wu2023latent}. For vertical instability, the challenge is to keep mesh resolved information without paying full attention cost over every grid point. A learned token representation is well matched to this structure: local mesh features can be compressed into a smaller set of low dimensional summaries, electromagnetic matrix embeddings can carry plasma to wall coupling information, and the result can be broadcast back to the mesh for spatial reconstruction.

This work introduces a Physics Attention Transformer (PAT) surrogate for vertical instability growth rate prediction. The architecture, summarized in Figure~\ref{fig:architecture}, maps an EFIT based free boundary equilibrium, scalar plasma parameters, active coil currents, passive structure information, and precomputed electromagnetic matrices to two outputs: the dominant MEQ-FGE-L growth rate $\hat{\gamma}$ and the associated perturbed toroidal current density $\widehat{\delta j}_\phi(R,Z)$. The attention block uses learned mesh summaries together with electromagnetic matrix embeddings; these tokens are a compact computational representation, not conserved quantities or guaranteed physical modes. A secondary surrogate for $\boldsymbol{\Xi}_{I_y}$ removes the expensive perturbed Grad-Shafranov response solve from deployment time inference.

The training corpus combines 10{,}162 C-Mod experimental equilibria with 8{,}734 synthetic SPARC scenarios. All labels are produced by MEQ-FGE-L, so the reported errors measure surrogate to solver fidelity rather than direct validation against nonlinear VDE dynamics or future SPARC experiments. Test splits are performed at the discharge or scenario trajectory level to reduce temporal leakage. On held out data, PAT reaches MAE $=5.4$~s$^{-1}$ on C-Mod and $12.7$~s$^{-1}$ on the full SPARC synthetic test set, while reconstructing spatial eigenfunctions with relative errors near 5\%. 

Rest of the paper proceeds as follows. Section~\ref{Sec.theory and model} describes the data and the linearized stability target. Section~\ref{Sec.MLVSO_impl} presents PAT, its loss function, and the $\Xi_{I_y}$ surrogate used for deployment. Section~\ref{sec:training} reports the training protocol and latency. Sections~\ref{sec:growth_rate_performance} and \ref{sec:cross_device} evaluates growth rate prediction, spatial reconstruction, baselines, uncertainty, interpretability, and transfer from C-Mod to SPARC. Section~\ref{sec:failure_analysis} examines operational  modes. Section~\ref{sec:deployment_implications} discusses implications for real-time use. Appendix~\ref{appendixA} documents the MLP baseline

\section{Data Sources and Linearized Stability Target}\label{Sec.theory and model}

Two data sources anchor the study. The experimental portion comes from C-Mod discharges recorded between 2012 and 2016 across Ohmic, L-mode, I-mode, and H-mode regimes\cite{Marmar_2015,Montes_2019}. For each discharge, diagnostics are sampled at 20~kHz and interpolated to 1~kHz time slices. Each slice contains an EFIT magnetic equilibrium, the vertical position of the plasma current centroid $z_{\mathrm{Ip}}$ from magnetics, wall current reconstructions from MEQ, and active coil currents. In the C-Mod diagnostic archive, this centroid channel is also denoted $z_{cc}$. MEQ-FGE-L then provides the instantaneous target growth rate $\gamma_{\text{MEQ}}$.

To avoid labeling the model with strongly off normal terminal behavior, we remove slices within 5~ms of last closed flux surface loss. The retained data still cover the 0 to 2~s pulse interval but exclude the final disruption phase, where topology changes and halo current physics fall outside the linearized target used here.
%#Each equilibrium state is labeled with vertical growth rate, $  \gamma_{MEQ}$, computed using high fidelity plasma coil conductor response model MEQ-FGE-L model. 

The SPARC Synthetic Library (SSL) extends the training domain into reactor relevant shapes that C-Mod cannot supply experimentally. It has two components:
\begin{enumerate}
    \item Reference pulse scenarios: four families of current and shape trajectories (1400, 1500, 1600, and 1700 series) generated by coupling MEQ-FGE and FGE-L with a passive conductor model of the SPARC vacuum vessel and stabilization coils\cite{wai2025}. The reference discharge duration is 30~s.
    \item Latin hypercube sampling: 8{,}734 additional synthetic scenarios generated by sampling plasma current $I_p$, safety factor $q_{95}$, elongation $\kappa$, and triangularity $\delta_{\mathrm{tri}}$ over prescribed operating ranges at fixed on axis toroidal field $B_{\mathrm{tor},0}$.
\end{enumerate}
\par
Both datasets are split into training (75\%), validation (10\%), and test (15\%) partitions. The split is stratified by machine and scenario type, then applied at the discharge or trajectory level. No discharge contributes slices to more than one partition. The validation set is used for hyperparameter selection; final metrics are reported only on the held out test sets.

\section{Linearized MEQ-FGE-L Target for \texorpdfstring{$\hat{\gamma}$}{gamma}}
\label{sec:IIB}

The MEQ-FGE-L label used for supervised learning is the dominant eigenvalue of a linearized plasma to conductor system. MEQ-FGE-L first computes the instantaneous nonrigid plasma response around a reconstructed equilibrium. That response is then folded into the external circuit dynamics, yielding a state matrix whose largest real eigenvalue defines the growth rate.

The notation in this section distinguishes scalars, vectors, and matrices explicitly. Scalars are written in ordinary italic type, vectors in bold lowercase or bold physical symbols, and matrices or linear operators in bold uppercase or bold Greek symbols. Continuous scalar fields such as $\delta j_\phi(R,Z)$ are not bold, while their discretized mesh vectors are bold.

The construction starts from small deviations about equilibrium values:
\begin{itemize}[noitemsep]
    \item $\delta\mathbf{I}_e(t) \in \mathbb{R}^{n_e}$: currents in external conducting filaments, comprising active coils and passive vessel structures, with $n_e=n_a+n_v$;
    \item $\delta\mathbf{I}_y(t) \in \mathbb{R}^{n_y}$: currents in the discretized plasma current basis;
    \item $\delta\mathbf{a}_g(t) \in \mathbb{R}^{n_g}$: geometric parameters describing the plasma shape (vertical position, elongation, etc.);
    \item $\delta\mathbf{V}_{\mathrm{ext}}(t) \in \mathbb{R}^{n_a}$: voltages applied to active control coils;
    \item $\delta\boldsymbol{\eta}(t) \in \mathbb{R}^{n_u}$: auxiliary control parameters, such as gas puffing rate or heating power.
\end{itemize}

On the slow timescale characteristic of vertical instability, the plasma maintains continuous force balance. This physical constraint is captured by the algebraic relation:
\begin{eqnarray}
    \mathcal{R}(\mathbf{I}_y, \mathbf{a}_g, \mathbf{I}_e, \boldsymbol{\eta}) = \mathbf{0}, \label{eq:force_balance}
\end{eqnarray}
where $\mathcal{R}: \mathbb{R}^{n_y + n_g + n_e + n_u} \to \mathbb{R}^{n_y + n_g}$ is the vector residual of the discretized equilibrium equations. The Grad-Shafranov operator in the plasma region $\Omega_p$ takes the sign convention:
\begin{equation*}
    \mathcal{G}[\psi] := \Delta^* \psi + \mu_0 R^2 p'(\psi) + FF'(\psi) = 0,
\end{equation*}
where the elliptic operator is defined as:
\begin{equation*}
    \Delta^* := R \frac{\partial}{\partial R}\left(\frac{1}{R}\frac{\partial}{\partial R}\right) + \frac{\partial^2}{\partial Z^2} = \frac{\partial^2}{\partial R^2}- \frac{1}{R}\frac{\partial}{\partial R} + \frac{\partial^2}{\partial Z^2}.
\end{equation*}
Here $R$ is major radius, $Z$ is vertical position, $\psi(R,Z)$ is the poloidal flux, $p(\psi)$ is the pressure profile, and $F(\psi)=RB_\phi$ is the poloidal current function. Primes denote differentiation with respect to $\psi$.

To analyze stability, we linearize the constraint~\eqref{eq:force_balance} about the equilibrium:
\begin{eqnarray}
    \mathbf{J}_{I_y}\delta\mathbf{I}_y + \mathbf{J}_{a_g}\delta\mathbf{a}_g + \mathbf{J}_e\delta\mathbf{I}_e + \mathbf{J}_{\eta}\delta\boldsymbol{\eta} = \mathbf{0}. \label{eq:linearized_constraint}
\end{eqnarray}
Here $\mathbf{J}_{I_y}:=\partial\mathcal{R}/\partial\mathbf{I}_y$, $\mathbf{J}_{a_g}:=\partial\mathcal{R}/\partial\mathbf{a}_g$, $\mathbf{J}_e:=\partial\mathcal{R}/\partial\mathbf{I}_e$, and $\mathbf{J}_{\eta}:=\partial\mathcal{R}/\partial\boldsymbol{\eta}$ are Jacobian matrices evaluated at the equilibrium. Define the plasma state vector and its Jacobian block as:
\begin{eqnarray}
    \delta\mathbf{x}_p := \begin{bmatrix} \delta\mathbf{I}_y \\ \delta\mathbf{a}_g \end{bmatrix} \in \mathbb{R}^{n_p}, \quad
    \mathbf{J}_p := \begin{bmatrix} \mathbf{J}_{I_y} & \mathbf{J}_{a_g} \end{bmatrix} \in \mathbb{R}^{n_p \times n_p},\label{eq:plasma_state}  
\end{eqnarray} where $n_p=n_y+n_g$. 
If $\mathbf{J}_p$ is nonsingular, or regularized in the same manner as the MEQ-FGE-L solve, the plasma response is:
\begin{eqnarray}
    \delta\mathbf{x}_p &=& \boldsymbol{\Xi} \begin{bmatrix} \delta\mathbf{I}_e \\ \delta\boldsymbol{\eta} \end{bmatrix}, \nonumber\\
    \boldsymbol{\Xi} &:=& -\mathbf{J}_p^{-1} \begin{bmatrix} \mathbf{J}_e & \mathbf{J}_{\eta} \end{bmatrix} \in \mathbb{R}^{n_p \times (n_e + n_u)}. \label{eq:response_operator}
\end{eqnarray}
The response operator $\boldsymbol{\Xi}$ captures the instantaneous, ideal response of the plasma current distribution and shape to changes in external conductor currents and auxiliary controls.

Since the plasma state $\delta\mathbf{x}_p$ concatenates current and geometric subvectors, we partition $\boldsymbol{\Xi}$ by rows. Let $\boldsymbol{\Xi}_e \in \mathbb{R}^{n_p \times n_e}$ and $\boldsymbol{\Xi}_{\eta} \in \mathbb{R}^{n_p \times n_u}$ denote the column blocks of $\boldsymbol{\Xi}$ corresponding to external currents and auxiliary controls, respectively. We further partition $\boldsymbol{\Xi}_e$ by rows according to the plasma state structure:
\begin{eqnarray}
    \boldsymbol{\Xi}_e = \begin{bmatrix} \boldsymbol{\Xi}_{I_y} \\ \boldsymbol{\Xi}_{a_g} \end{bmatrix}, \quad \boldsymbol{\Xi}_{I_y} \in \mathbb{R}^{n_y \times n_e}, \quad \boldsymbol{\Xi}_{a_g} \in \mathbb{R}^{n_g \times n_e}, \label{eq:Xi_partition}
\end{eqnarray}
where $\boldsymbol{\Xi}_{I_y}$ maps external current perturbations to plasma current perturbations, and $\boldsymbol{\Xi}_{a_g}$ maps them to geometric shape perturbations. An analogous partition applies to $\boldsymbol{\Xi}_{\eta}$, yielding subblocks $\boldsymbol{\Xi}_{I_y,\eta} \in \mathbb{R}^{n_y \times n_u}$ and $\boldsymbol{\Xi}_{a_g,\eta} \in \mathbb{R}^{n_g \times n_u}$.

The instantaneous plasma response enters the external circuit through the linearized Faraday to Ohm equations:
\begin{eqnarray}
    \mathbf{M}_{ee}\delta \dot{\mathbf{I}}_e + \mathbf{M}_{ey}\delta \dot{\mathbf{I}}_y + \mathbf{R}_e \delta\mathbf{I}_e = \mathbf{B}_{ea}\delta\mathbf{V}_{\mathrm{ext}}, \label{eq:circuit}
\end{eqnarray}
where $\mathbf{M}_{ee} \in \mathbb{R}^{n_e \times n_e}$ is the mutual inductance matrix among external conductors, $\mathbf{M}_{ey} \in \mathbb{R}^{n_e \times n_y}$ is the mutual inductance matrix coupling plasma current basis functions to external conductors, $\mathbf{R}_e \in \mathbb{R}^{n_e \times n_e}$ is the diagonal resistance matrix, and $\mathbf{B}_{ea} \in \{0,1\}^{n_e \times n_a}$ selects the active coils.

From~\eqref{eq:response_operator} and~\eqref{eq:Xi_partition}, the plasma current response to external current perturbations is $\delta\mathbf{I}_y = \boldsymbol{\Xi}_{I_y} \delta\mathbf{I}_e + \boldsymbol{\Xi}_{I_y,\eta} \delta\boldsymbol{\eta}$. The quasi static approximation treats $\boldsymbol{\Xi}$ as fixed over one wall current timescale. This is justified when the plasma equilibrates on the Alfv\'{e}n timescale ($\sim\!\mu$s), much faster than resistive wall current evolution ($\sim$ms), so $\dot{\boldsymbol{\Xi}} \approx \mathbf{0}$ and:
\begin{eqnarray}
    \delta \dot{\mathbf{I}}_y = \boldsymbol{\Xi}_{I_y}\,\delta \dot{\mathbf{I}}_e + \boldsymbol{\Xi}_{I_y,\eta}\,\dot{\delta\boldsymbol{\eta}}. \label{eq:Iy_dot}
\end{eqnarray}
Substituting~\eqref{eq:Iy_dot} into~\eqref{eq:circuit} yields a closed form equation for the external currents:
\begin{eqnarray}
    \mathbf{L}_{ee}^* \delta \dot{\mathbf{I}}_e + \mathbf{R}_e \delta\mathbf{I}_e = \mathbf{B}_{ea}\delta\mathbf{V}_{\mathrm{ext}}- \mathbf{M}_{ey}\boldsymbol{\Xi}_{I_y,\eta}\, \dot{\delta\boldsymbol{\eta}}, \label{eq:effective_circuit}
\end{eqnarray}
where the effective inductance matrix is:
\begin{eqnarray}
    \mathbf{L}_{ee}^* := \mathbf{M}_{ee} + \mathbf{M}_{ey}\boldsymbol{\Xi}_{I_y} \in \mathbb{R}^{n_e \times n_e}. \label{eq:Lstar}
\end{eqnarray}
The sign convention in Eq.~\ref{eq:Lstar} follows from the definition of $\boldsymbol{\Xi}$ in Eq.~\ref{eq:response_operator}; the minus sign from the linearized equilibrium solve is already contained in $\boldsymbol{\Xi}_{I_y}$. Stability is determined by the spectrum of $-(\mathbf{L}_{ee}^*)^{-1}\mathbf{R}_e$, not by entrywise signs in $\boldsymbol{\Xi}_{I_y}$. The matrix $\mathbf{L}_{ee}^*$ is the external circuit inductance modified by nonrigid plasma response.

With the effective circuit defined, vertical stability reduces to an eigenvalue problem. For stability analysis, we consider the autonomous system with $\delta\mathbf{V}_{\mathrm{ext}} = \mathbf{0}$ and $\dot{\delta\boldsymbol{\eta}} = \mathbf{0}$:
\begin{eqnarray}
    \delta \dot{\mathbf{I}}_e = \mathbf{A} \delta\mathbf{I}_e, \quad \text{where} \quad \mathbf{A} :=-(\mathbf{L}_{ee}^*)^{-1}\mathbf{R}_e \in \mathbb{R}^{n_e \times n_e}. \label{eq:state_matrix}
\end{eqnarray}
The MEQ-FGE-L label is the largest real part among the eigenvalues of $\mathbf{A}$:
\begin{eqnarray}
    \gamma_{\mathrm{MEQ}} := \max_{k=1,\ldots,n_e} \operatorname{Re}(\lambda_k), \quad \text{where} \quad \det(\mathbf{A}- \lambda_k \mathbf{I}_{n_e}) = 0. \label{eq:growth_rate}
\end{eqnarray}
A positive $\gamma_{\mathrm{MEQ}}$ indicates an unstable mode that grows exponentially in the linearized model. Although $\mathbf{A}$ is not guaranteed to be symmetric, the dominant vertical eigenvalue in the MEQ-FGE-L cases studied here is real to numerical tolerance. If a complex pair is encountered, the physical perturbation is obtained from the real part after fixing the eigenvector phase.

The spatial structure of the unstable mode is reconstructed from the eigenvector $\mathbf{v} \in \mathbb{C}^{n_e}$ associated with the dominant eigenvalue. The corresponding plasma current perturbation $\delta\mathbf{I}_y = \boldsymbol{\Xi}_{I_y} \mathbf{v}$ yields the 2D perturbed toroidal current density:
\begin{eqnarray}
    \delta j_\phi(R, Z) = \operatorname{Re}\!\left[\sum_{k=1}^{n_y} (\boldsymbol{\Xi}_{I_y}\, \mathbf{v})_k \,\varphi_k(R, Z)\right]. \label{eq:eigenfunction}
\end{eqnarray}
where $\{\varphi_k(R,Z)\}_{k=1}^{n_y}$ are current density basis functions on the plasma mesh, normalized so that a coefficient vector with units of current reconstructs a field with units A/m$^2$ after the cell area normalization used by MEQ-FGE-L. Eigenvectors are normalized to unit $L^2$ norm, and their phase is fixed by requiring the largest magnitude component of $\mathbf{v}$ to be real and positive. The supervised eigenvector label $\mathbf{v}_{\mathrm{MEQ}}$ denotes this phase fixed real representative. The same convention is applied before supervising $\delta j_\phi$.

%%%%%%%%%%%%%%%%%%%%%%%%%%%%%%%%%%%%%%%%%%%%%%%%%%%%%%%%%%%%%%%%%%%%%%%%
\section{Physics Attention Surrogate}\label{Sec.MLVSO_impl}

PAT maps an equilibrium representation to the scalar growth rate and to the mode structure that accompanies it. Let $f_{\boldsymbol{\theta}}: \mathcal{X} \to \mathbb{R} \times \mathbb{R}^{n_e} \times \mathbb{R}^{N_m}$ denote the surrogate, where $\mathcal{X}$ is the space of equilibrium representations and $N_m$ is the number of PAT mesh points. When the MEQ-FGE-L current basis and PAT mesh differ, the solver field is evaluated on the PAT mesh through the basis expansion in Eq.~\ref{eq:eigenfunction}; in the present datasets this sampled field has dimension $N_m$. The three outputs are the growth rate $\hat{\gamma}_{\boldsymbol{\theta}}$, a conductor space eigenvector $\hat{\mathbf{v}}_{\boldsymbol{\theta}} \in \mathbb{R}^{n_e}$, and a mesh resolved field vector $\widehat{\boldsymbol{\delta j}}_{\phi,\boldsymbol{\theta}} \in \mathbb{R}^{N_m}$.

\label{sec:IIIA}

The inputs follow the same physical decomposition as the outputs: local mesh fields are paired with global scalars, coil currents, and electromagnetic operators. Let $N_m \approx 98{,}304$ denote the number of PAT mesh points, $M = 64$ the number of learned mesh tokens, and $C = 256$ the hidden dimension. For each grid point $\mathbf{r}_i = (R_i, Z_i)^T$, the raw feature vector $\mathbf{u}_i \in \mathbb{R}^{C_0}$ concatenates geometric, equilibrium, global, and coil current information:
\begin{widetext}
\begin{eqnarray}
    \mathbf{u}_i = \left[\underbrace{\frac{R_i- R_0}{a}, \frac{Z_i- Z_0}{a}, R_i^{-1}}_{\text{geometry}}, \underbrace{\psi_0(\mathbf{r}_i), \partial_R\psi_0(\mathbf{r}_i), \ldots, |\mathbf{B}_{\mathrm{eq}}(\mathbf{r}_i)|}_{\text{equilibrium fields}}, \underbrace{I_p, q_{95}, \ldots, \beta_p}_{\text{global parameters}}, \underbrace{\mathbf{I}_{\mathrm{coil}} \in \mathbb{R}^{n_{\mathrm{coil}}}}_{\text{coil currents}}\right]^T, \label{eq:input_features}
\end{eqnarray}
\end{widetext}
where $R_0$ and $Z_0$ are the equilibrium magnetic axis coordinates used for normalization, $a$ is the minor radius normalization, $\psi_0$ is the equilibrium poloidal flux, $\mathbf{B}_{\mathrm{eq}}$ is the equilibrium magnetic field, $\mathbf{I}_{\mathrm{coil}}$ is the vector of active coil current inputs, $n_{\mathrm{coil}}$ is its dimension, and $C_0$ is the raw feature dimension. The ellipses denote additional EFIT derived field channels or scalar equilibrium parameters in the same category.

The raw vector is projected into the model's hidden space:
\begin{eqnarray}
    \mathbf{x}_i^{(0)} = \mathbf{W}_e\, \mathbf{u}_i + \mathbf{b}_e + \mathbf{P}(\mathbf{r}_i), \label{eq:embedding}
\end{eqnarray}
where $\mathbf{W}_e \in \mathbb{R}^{C \times C_0}$ and $\mathbf{b}_e \in \mathbb{R}^C$ are learnable parameters, and $\mathbf{P}(\mathbf{r}_i) \in \mathbb{R}^C$ provides rotary positional encoding of the spatial coordinates.

Four matrix tokens carry the electromagnetic quantities that are not local mesh fields: $\mathbf{M}_{ee} \in \mathbb{R}^{n_e \times n_e}$, $\mathbf{M}_{ey} \in \mathbb{R}^{n_e \times n_y}$, $\mathbf{R}_e \in \mathbb{R}^{n_e \times n_e}$, and the plasma current response suboperator $\boldsymbol{\Xi}_{I_y} \in \mathbb{R}^{n_y \times n_e}$ (Eq.~\ref{eq:Xi_partition}). The $\mathbf{M}_{ey}$ token is physically motivated because the effective inductance $\mathbf{L}_{ee}^* = \mathbf{M}_{ee} + \mathbf{M}_{ey}\boldsymbol{\Xi}_{I_y}$ (Eq.~\ref{eq:Lstar}) depends on plasma to conductor coupling directly. Without such a pathway, the network would have to infer this coupling from geometry alone.

The matrices $\mathbf{M}_{ee}$, $\mathbf{M}_{ey}$, and $\mathbf{R}_e$ depend only on conductor and mesh geometry; for a given machine they are constants, precomputed once. $\boldsymbol{\Xi}_{I_y}$ is different: it requires solving the perturbed Grad-Shafranov equation at each equilibrium, which dominates the ${\sim}3$-min wall clock time of MEQ-FGE-L. To remove this bottleneck at inference, we train a secondary surrogate $\widehat{\boldsymbol{\Xi}}_{I_y}$, described below, that predicts $\boldsymbol{\Xi}_{I_y}$ from the same EFIT features available in real time, eliminating the need for the expensive solver in the deployment loop.

Each matrix is flattened, projected to dimension $C$ via a learned linear layer, and appended to the token set at every attention layer.

\label{sec:IIIB}

\begin{figure*}[htbp]
\centering
\includegraphics[width=0.98\textwidth]{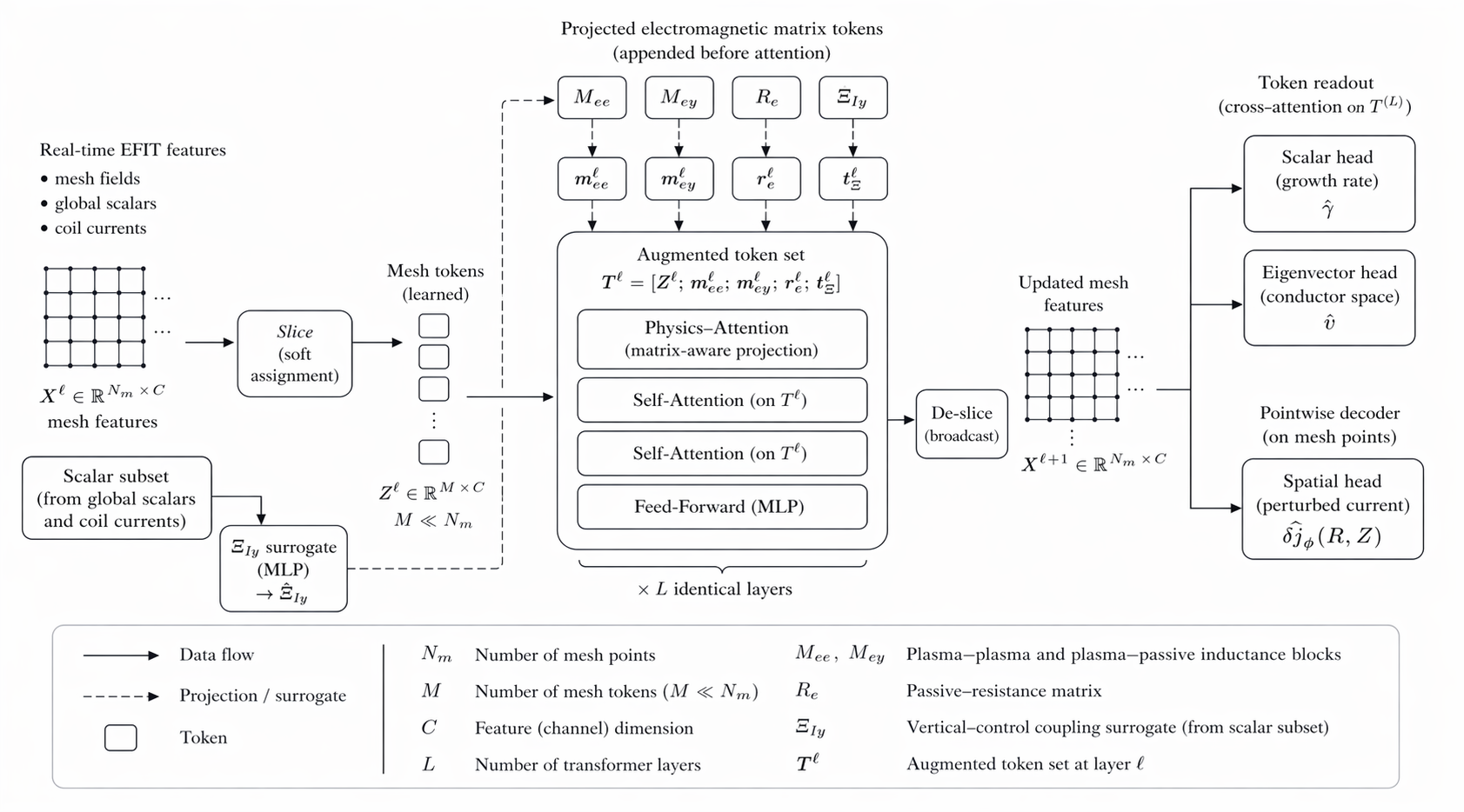}
\caption{Physics Attention Transformer schematic. Real time EFIT-LIKE features provide mesh fields, global scalars, and coil currents; the $\boldsymbol{\Xi}_{I_y}$ surrogate maps the scalar subset to $\widehat{\boldsymbol{\Xi}}_{I_y}$. Mesh features $\mathbf{X}^\ell \in \mathbb{R}^{N_m \times C}$ are softly sliced into $M$ learned mesh tokens, while projected matrix tokens for $\mathbf{M}_{ee}$, $\mathbf{M}_{ey}$, $\mathbf{R}_e$, and $\widehat{\boldsymbol{\Xi}}_{I_y}$ are appended before token self attention. De slicing broadcasts only the mesh token outputs back to the grid. After $L=6$ PAT blocks, cross attention readout predicts $\hat{\gamma}$ and $\hat{\mathbf{v}}$, while a pointwise decoder predicts $\widehat{\delta j}_\phi(R,Z)$.}
\label{fig:architecture}
\end{figure*}

The core architecture comprises $L = 6$ identical attention blocks. Each block $\ell$ transforms mesh representations $\{\mathbf{x}_i^{\ell}\}_{i=1}^{N_m}$ by compressing the mesh into learned tokens, allowing those tokens and the electromagnetic embeddings to exchange global information, and broadcasting the attended mesh token information back to the grid before a feed forward update. Figure~\ref{fig:architecture} summarizes the full path from EFIT inputs to the growth rate, eigenvector, and spatial field outputs.

Information from $N_m$ mesh points is condensed into $M$ learned mesh tokens via soft assignment. A two layer MLP computes assignment logits:
\begin{eqnarray}
    \mathbf{S}^{\ell} = \GELU\!\left(\mathbf{X}^{\ell} {\mathbf{W}_s^{\ell}}^T\right) {\mathbf{W}_{\mathrm{slice}}^{\ell}}^T \in \mathbb{R}^{N_m \times M}, \label{eq:slice_logits}
\end{eqnarray}
where $\mathbf{X}^{\ell} \in \mathbb{R}^{N_m \times C}$ stacks all mesh representations, $\mathbf{W}_s^{\ell} \in \mathbb{R}^{d_h \times C}$ projects to a hidden dimension $d_h$, and $\mathbf{W}_{\mathrm{slice}}^{\ell} \in \mathbb{R}^{M \times d_h}$ produces $M$ logits per point. The assignment weights are obtained via row wise softmax:
\begin{eqnarray}
    w_{ij}^{\ell} = \frac{\exp(\mathbf{S}_{ij}^{\ell})}{\sum_{k=1}^{M}\exp(\mathbf{S}_{ik}^{\ell})}, \quad \sum_{j=1}^{M} w_{ij}^{\ell} = 1 \quad \forall\, i. \label{eq:soft_assignment}
\end{eqnarray}
Each mesh point distributes its representation across all $M$ tokens; column wise normalization is not applied, so token sizes are determined by the learned assignment structure rather than being uniform.

Mesh tokens are formed as normalized weighted averages:
\begin{eqnarray}
    \mathbf{z}_j^{\ell} = \frac{\sum_{i=1}^{N_m} w_{ij}^{\ell} \mathbf{x}_i^{\ell}}{\sum_{i=1}^{N_m} w_{ij}^{\ell}} \in \mathbb{R}^C. \label{eq:token_formation}
\end{eqnarray}
The denominator normalizes by the effective ``soft count'' of points assigned to token $j$, ensuring consistent token magnitudes regardless of region size.

The mesh derived tokens are stacked as $\mathbf{Z}^{\ell}\in\mathbb{R}^{M\times C}$. Four projected electromagnetic tokens, denoted $\mathbf{m}_{ee}^{\ell}$, $\mathbf{m}_{ey}^{\ell}$, $\mathbf{r}_e^{\ell}$, and $\mathbf{t}_{\Xi}^{\ell}$, are appended to form:
\begin{eqnarray}
    \mathbf{T}^{\ell} = \left[\mathbf{Z}^{\ell};\,\mathbf{m}_{ee}^{\ell};\,\mathbf{m}_{ey}^{\ell};\,\mathbf{r}_e^{\ell};\,\mathbf{t}_{\Xi}^{\ell}\right] \in \mathbb{R}^{(M+4)\times C}. \label{eq:augmented_tokens}
\end{eqnarray}
Self attention is applied to this augmented token set:
\begin{eqnarray}
    \tilde{\mathbf{T}}^{\ell} = \operatorname{softmax}\!\left(\frac{\mathbf{Q}\mathbf{K}^T}{\sqrt{C_h}}\right)\mathbf{V} \in \mathbb{R}^{(M+4) \times C}, \label{eq:self_attention}
\end{eqnarray}
where $\mathbf{Q} = \mathbf{T}^{\ell}\mathbf{W}_Q$, $\mathbf{K} = \mathbf{T}^{\ell}\mathbf{W}_K$, and $\mathbf{V} = \mathbf{T}^{\ell}\mathbf{W}_V$ are the query, key, and value projections with $\mathbf{W}_Q, \mathbf{W}_K, \mathbf{W}_V \in \mathbb{R}^{C \times C}$. The per head dimension is $C_h = C/H$ for $H$ attention heads. Only the first $M$ rows of $\tilde{\mathbf{T}}^{\ell}$, denoted $\tilde{\mathbf{Z}}_{\mathrm{mesh}}^{\ell}$, are de sliced back to the mesh.

Attended token information is broadcast back to mesh points using the same assignment weights:
\begin{eqnarray}
    \operatorname{Deslice}(\tilde{\mathbf{Z}}_{\mathrm{mesh}}^{\ell})_i = \sum_{j=1}^{M} w_{ij}^{\ell} \tilde{\mathbf{z}}_{\mathrm{mesh},j}^{\ell} \in \mathbb{R}^C. \label{eq:deslice}
\end{eqnarray}

The full block employs prenormalization with residual connections:
\begin{eqnarray}
    &\mathbf{x}_i'& = \mathbf{x}_i^{\ell} + \operatorname{Deslice}\!\left(\operatorname{Attn}\!\left(\operatorname{Tokenize}\!\left(\operatorname{LN}(\mathbf{X}^{\ell})\right)\right)\right)_i, \label{eq:block_attn} \\
    &\mathbf{x}_i^{\ell+1}& = \mathbf{x}_i' + \operatorname{FFN}\!\left(\operatorname{LN}(\mathbf{x}_i')\right), \label{eq:block_ffn}
\end{eqnarray}
where LN denotes layer normalization and FFN is a two layer MLP with GELU activation and expansion factor 4.

\label{sec:IIIC}

After the final PAT block, dedicated heads produce scalar, conductor space, and mesh resolved outputs. A cross attention readout extracts the growth rate from the final augmented token set $\mathbf{T}^{(L)} \in \mathbb{R}^{(M+4) \times C}$. Learnable query tokens $\mathbf{Q}_0 \in \mathbb{R}^{M_q \times C}$ attend to the augmented token set:
\begin{eqnarray}
    \mathbf{Q}_1 = \operatorname{softmax}\!\left(\frac{\mathbf{Q}_0 \mathbf{W}_Q' (\mathbf{T}^{(L)}\mathbf{W}_K')^T}{\sqrt{C_h}}\right) \mathbf{T}^{(L)}\mathbf{W}_V' \in \mathbb{R}^{M_q \times C}. \label{eq:cross_attention}
\end{eqnarray}
The growth rate is predicted by pooling and passing through an MLP:
\begin{eqnarray}
    \hat{\gamma} = \operatorname{MLP}_\gamma\!\left(\frac{1}{M_q}\sum_{k=1}^{M_q}[\mathbf{Q}_1]_k\right), \label{eq:gamma_head}
\end{eqnarray}
where $[\mathbf{Q}_1]_k$ denotes the $k$-th row of $\mathbf{Q}_1$.

A second readout head maps the pooled query representation to an eigenvector in conductor current space:
\begin{eqnarray}
    \hat{\mathbf{v}} = \operatorname{MLP}_v\!\left(\frac{1}{M_q}\sum_{k=1}^{M_q}[\mathbf{Q}_1]_k\right) \in \mathbb{R}^{n_e}, \label{eq:eigvec_head}
\end{eqnarray}
where $\operatorname{MLP}_v: \mathbb{R}^C \to \mathbb{R}^{n_e}$ is a two layer MLP with hidden dimension $2n_e$. This head shares the cross attention features $\mathbf{Q}_1$ with the growth rate head but employs separate MLP parameters. The predicted eigenvector enters the eigenvalue consistency loss (Eq.~\ref{eq:loss_eigen}) and normalization constraint (Eq.~\ref{eq:loss_norm}), enforcing that $(\hat{\gamma}, \hat{\mathbf{v}})$ jointly approximate the dominant eigenpair of the state matrix $\mathbf{A}$.

The perturbed toroidal current density $\hat{\delta j}_\phi(R,Z)$ is predicted directly from final mesh representations via a learned linear projection:
\begin{eqnarray}
    \widehat{\delta j}_\phi(\mathbf{r}_i) = \mathbf{w}_\phi^T \mathbf{x}_i^{(L)} + b_\phi, \label{eq:spatial_head}
\end{eqnarray}
where $\mathbf{w}_\phi \in \mathbb{R}^C$ and $b_\phi \in \mathbb{R}$ are learnable parameters. This spatial output is supervised by the spatial field fidelity loss (Eq.~\ref{eq:loss_spatial}) and regularized by the boundary condition loss (Eq.~\ref{eq:loss_bc}). The eigenvector and spatial heads are separately supervised outputs that share the same latent representation; the eigenvalue consistency loss constrains $\hat{\mathbf{v}}$, while the spatial loss constrains $\widehat{\delta j}_\phi$ directly.

\label{sec:IIC}

With these heads defined, training can be written as a composite objective rather than a collection of disconnected penalties. For a labeled dataset $\mathcal{D} = \{(\boldsymbol{\xi}^{(i)}, \gamma^{(i)}_{\mathrm{MEQ}}, \mathbf{v}^{(i)}_{\mathrm{MEQ}}, \boldsymbol{\delta j}^{(i)}_{\phi,\mathrm{MEQ}})\}_{i=1}^{N_{\text{data}}}$, where $\boldsymbol{\xi}^{(i)}$ is the full equilibrium input representation and the spatial perturbation follows Eq.~\ref{eq:eigenfunction}, PAT minimizes:
\begin{eqnarray}
    \mathcal{L}(\boldsymbol{\theta}) = \lambda_d \mathcal{L}_{\text{data}} + \lambda_s \mathcal{L}_{\text{spatial}} + \lambda_e \mathcal{L}_{\text{eigen}} + \lambda_b \mathcal{L}_{\text{bc}} \nonumber\\ + \lambda_n \mathcal{L}_{\text{norm}} + \lambda_r \mathcal{L}_{\text{reg}}, \label{eq:total_loss}
\end{eqnarray}
where the hyperparameters $\{\lambda_d, \lambda_s, \lambda_e, \lambda_b, \lambda_n, \lambda_r\}$ are initialized to balance loss components and subsequently tuned during optimization.

The scalar term matches predicted growth rates to MEQ-FGE-L labels:
\begin{eqnarray}
    \mathcal{L}_{\text{data}}(\boldsymbol{\theta}) = \frac{1}{N_{\text{data}}}\sum_{i=1}^{N_{\text{data}}} \left(\hat{\gamma}_{\boldsymbol{\theta}}(\boldsymbol{\xi}^{(i)})- \gamma^{(i)}_{\mathrm{MEQ}}\right)^2. \label{eq:loss_data}
\end{eqnarray}
The spatial term compares the decoded current density field with the MEQ-FGE-L perturbation:
\begin{eqnarray}
    \mathcal{L}_{\text{spatial}}(\boldsymbol{\theta}) = \frac{1}{N_{\text{data}}}\sum_{i=1}^{N_{\text{data}}} \frac{\|\widehat{\boldsymbol{\delta j}}^{(i)}_{\phi,\boldsymbol{\theta}}- \boldsymbol{\delta j}^{(i)}_{\phi,\mathrm{MEQ}}\|_{\Omega_p,h}^2}{\|\boldsymbol{\delta j}^{(i)}_{\phi,\mathrm{MEQ}}\|_{\Omega_p,h}^2 + \varepsilon_{\mathrm{num}}}, \label{eq:loss_spatial}
\end{eqnarray}
where $\|\mathbf{y}\|_{\Omega_p,h}^2 := \sum_{m=1}^{N_m} |y_m|^2 \Delta A_m$ is the area weighted mesh norm, $\Delta A_m$ is the cell area weight at mesh point $m$, and $\varepsilon_{\mathrm{num}}>0$ prevents division by zero for marginally stable cases with vanishing perturbation. The denominator normalizes by the solver field magnitude to weight equilibria comparably regardless of absolute perturbation amplitude.

The eigenvalue term ties the scalar and conductor space outputs back to Eq.~\ref{eq:state_matrix} to \ref{eq:growth_rate}:
\begin{eqnarray}
    \mathcal{L}_{\text{eigen}}(\boldsymbol{\theta}) = \frac{1}{N_{\text{data}}}\sum_{i=1}^{N_{\text{data}}} \frac{\|\mathbf{A}^{(i)} \hat{\mathbf{v}}^{(i)}_{\boldsymbol{\theta}}- \hat{\gamma}^{(i)}_{\boldsymbol{\theta},\mathrm{phys}} \hat{\mathbf{v}}^{(i)}_{\boldsymbol{\theta}}\|_2^2}{\|\hat{\mathbf{v}}^{(i)}_{\boldsymbol{\theta}}\|_2^2 + \varepsilon_{\mathrm{num}}}, \label{eq:loss_eigen}
\end{eqnarray}
where $\hat{\mathbf{v}}^{(i)}_{\boldsymbol{\theta}} \in \mathbb{R}^{n_e}$ is the network's predicted real eigenvector for equilibrium $i$, $\mathbf{A}^{(i)}$ is the state matrix computed from the equilibrium's inductance and resistance data, and $\hat{\gamma}^{(i)}_{\boldsymbol{\theta},\mathrm{phys}}$ is the de standardized growth rate prediction in s$^{-1}$. This regularizer encourages eigenpairs consistent with the linearized stability equations without requiring explicit eigendecomposition during training.

The boundary term is written in perturbed poloidal flux rather than current density. Since the network predicts $\widehat{\boldsymbol{\delta j}}_\phi$, we compute $\widehat{\delta\psi}_{\boldsymbol{\theta}}$ from $\widehat{\boldsymbol{\delta j}}_{\phi,\boldsymbol{\theta}}$ using the same axisymmetric Green's function map used in MEQ-FGE-L post processing. In the ideal wall limit, $\delta\psi|_{\partial\Omega_w}=0$, so:
\begin{eqnarray}
    \mathcal{L}_{\text{bc}}(\boldsymbol{\theta}) = \frac{1}{N_{\text{data}} \cdot N_b}\sum_{i=1}^{N_{\text{data}}}\sum_{k=1}^{N_b} \left|\widehat{\delta\psi}_{\boldsymbol{\theta}}(\boldsymbol{\xi}^{(i)}, \mathbf{r}_k^{\text{wall}})\right|^2, \label{eq:loss_bc}
\end{eqnarray}
where $\{\mathbf{r}_k^{\text{wall}}\}_{k=1}^{N_b}$ are collocation points on the wall boundary $\partial\Omega_w$, and $N_b$ is the number of such points. This term constrains the decoded field in an ideal wall regularization limit; it is not an independent validation of the resistive wall physics in MEQ-FGE-L.

The eigenvector head also needs a scale convention. To prevent the trivial solution $\hat{\mathbf{v}}_{\boldsymbol{\theta}} \equiv \mathbf{0}$ and ensure a well defined eigenvector, we enforce unit normalization:
\begin{eqnarray}
    \mathcal{L}_{\text{norm}}(\boldsymbol{\theta}) = \frac{1}{N_{\text{data}}}\sum_{i=1}^{N_{\text{data}}}\left(\|\hat{\mathbf{v}}^{(i)}_{\boldsymbol{\theta}}\|_2- 1\right)^2. \label{eq:loss_norm}
\end{eqnarray}
Standard $L_2$ weight decay completes the objective:
\begin{eqnarray}
    \mathcal{L}_{\text{reg}}(\boldsymbol{\theta}) = \|\boldsymbol{\theta}\|_2^2. \label{eq:loss_reg}
\end{eqnarray}

\label{sec:xi_surrogate}

The remaining deployment bottleneck is $\boldsymbol{\Xi}_{I_y}$. Computing this operator from the perturbed Grad-Shafranov equation takes ${\sim}3$~min per equilibrium, making it unavailable within a PCS control cycle. To close the deployment loop, we train a secondary surrogate $\widehat{\boldsymbol{\Xi}}_{I_y}$ that predicts the response operator from a real time scalar subset of the PAT inputs.

We note that $\boldsymbol{\Xi}_{I_y} \in \mathbb{R}^{n_y \times n_e}$ is a large matrix, but its response is concentrated in a small number of global modes. We therefore predict a low rank factorization $\widehat{\boldsymbol{\Xi}}_{I_y} = \mathbf{U}\mathbf{V}$, where $\mathbf{U} \in \mathbb{R}^{n_y \times r}$ and $\mathbf{V} \in \mathbb{R}^{r \times n_e}$ with $r \ll \min(n_y, n_e)$. The rank $r$ is the smallest value for which the mean retained Frobenius energy of the training set response matrices exceeds 99\%. A three layer MLP maps the global scalar vector $\mathbf{g}$ and coil current vector $\mathbf{I}_{\mathrm{coil}}$ to the entries of $\mathbf{U}$ and $\mathbf{V}$, where $\mathbf{g}$ contains $I_p$, $q_{95}$, $\beta_p$, $\kappa$, $\ell_i$, and triangularity components such as $\delta_u$ and $\delta_\ell$. The $\boldsymbol{\Xi}_{I_y}$ surrogate is trained on the training split and evaluated on the same held out partitions used for PAT.

At inference the chained pipeline is:
\begin{equation}
    \text{EFIT} \;\xrightarrow{}\; \widehat{\boldsymbol{\Xi}}_{I_y} \;\xrightarrow{\sim 1\,\text{ms}}\; \text{PAT} \;\xrightarrow{\sim 35\,\text{ms}}\; (\hat{\gamma},\, \widehat{\delta j}_\phi). \label{eq:pipeline}
\end{equation}
The full chain stays under 40~ms without an iterative Grad-Shafranov solve. In an end to end test, replacing the solver computed $\boldsymbol{\Xi}_{I_y}$ with $\widehat{\boldsymbol{\Xi}}_{I_y}$ changed growth rate and spatial field metrics within the displayed numerical precision of the reported tables on both test sets. This result is useful for deployment because it removes the circular dependency on the slow solver at inference. It does not validate the physical correctness of $\boldsymbol{\Xi}_{I_y}$ itself; that uncertainty remains part of the MEQ-FGE-L labeling assumption.

Finally, we summarise our ML architecture in Algorithm~\ref{alg:pat}, which is being used only during training.

\begin{algorithm}[htbp]
\caption{PAT inference path and training objective with $\boldsymbol{\Xi}_{I_y}$ surrogate}\label{alg:pat}
\begin{algorithmic}[1]
\REQUIRE Mesh points $\{\mathbf{r}_i=(R_i, Z_i)^T\}_{i=1}^{N_m}$, equilibrium fields, global parameter vector $\mathbf{g}$, coil current vector $\mathbf{I}_{\mathrm{coil}}$; precomputed $\mathbf{M}_{ee}$, $\mathbf{M}_{ey}$, $\mathbf{R}_e$; training labels $\gamma_{\mathrm{MEQ}}$, $\boldsymbol{\delta j}_{\phi,\mathrm{MEQ}}$, $\mathbf{A}$, and $\boldsymbol{\Xi}_{I_y}$ when optimizing losses
\ENSURE Growth rate $\hat{\gamma}$, eigenvector $\hat{\mathbf{v}}$, spatial field $\widehat{\delta j}_\phi$
\STATE \textit{  $\boldsymbol{\Xi}_{I_y}$ surrogate  }
\STATE $\mathbf{U}, \mathbf{V} \leftarrow \text{MLP}_{\Xi}(\mathbf{g},\, \mathbf{I}_{\mathrm{coil}})$ \hfill $\triangleright$ low rank factors
\STATE $\widehat{\boldsymbol{\Xi}}_{I_y} \leftarrow \mathbf{U}\mathbf{V}$ \hfill $\triangleright\; \mathbb{R}^{n_y \times n_e}$
\STATE \textit{ --- Embedding ---  }
\STATE $\mathbf{x}_i^{(0)} \leftarrow \mathbf{W}_e\, \mathbf{u}_i + \mathbf{b}_e + \mathbf{P}(\mathbf{r}_i)$ for each mesh point $i$
\STATE Project $\mathbf{M}_{ee},\, \mathbf{M}_{ey},\, \mathbf{R}_e,\, \widehat{\boldsymbol{\Xi}}_{I_y}$ to dimension $C$
\FOR{$\ell = 0$ to $L-1$}
    \STATE \textit{ --- Tokenization --- }
    \STATE $\mathbf{S}^\ell \leftarrow \GELU(\mathbf{X}^\ell {\mathbf{W}_s^\ell}^T)\,{\mathbf{W}_{\mathrm{slice}}^\ell}^T$ \hfill $\triangleright\; \mathbb{R}^{N_m \times M}$
    \STATE $w_{ij}^\ell \leftarrow \softmax_j(\mathbf{S}_{ij}^\ell)$ \hfill $\triangleright$ row wise
    \STATE $\mathbf{z}_j^\ell \leftarrow \sum_i w_{ij}^\ell \mathbf{x}_i^\ell \,/\, \sum_i w_{ij}^\ell$ \hfill $\triangleright$ $M$ learned mesh tokens
    \STATE Append matrix tokens to $\mathbf{Z}^\ell$ to form $\mathbf{T}^{\ell}$
    \STATE \textit{ --- Attention ---  }
    \STATE $\tilde{\mathbf{T}}^\ell \leftarrow \text{MultiHead}(\mathbf{T}^\ell \mathbf{W}_Q,\, \mathbf{T}^\ell \mathbf{W}_K,\, \mathbf{T}^\ell \mathbf{W}_V)$
    \STATE \textit{  De slice  }
    \STATE $\mathbf{x}_i' \leftarrow \mathbf{x}_i^\ell + \sum_{j=1}^{M} w_{ij}^\ell \tilde{\mathbf{z}}_{\mathrm{mesh},j}^\ell$ \hfill $\triangleright$ broadcast mesh token outputs only
    \STATE $\mathbf{x}_i^{\ell+1} \leftarrow \mathbf{x}_i' + \FFN(\LayerNorm(\mathbf{x}_i'))$
\ENDFOR
\STATE \textit{ --- Readout --- }
\STATE $\mathbf{Q}_1 \leftarrow \text{CrossAttn}(\mathbf{Q}_0,\, \mathbf{T}^{(L)})$ \hfill $\triangleright$ $M_q$ queries
\STATE $\hat{\gamma} \leftarrow \text{MLP}_\gamma(\text{pool}(\mathbf{Q}_1))$
\STATE $\hat{\mathbf{v}} \leftarrow \text{MLP}_v(\text{pool}(\mathbf{Q}_1))$ \hfill $\triangleright$ $\in \mathbb{R}^{n_e}$
\STATE $\widehat{\delta j}_\phi(\mathbf{r}_i) \leftarrow \mathbf{w}_\phi^T \mathbf{x}_i^{(L)} + b_\phi$ \hfill $\triangleright$ pointwise
\STATE \textit{  Training only loss  }
\STATE De standardize $\hat{\gamma}$ to $\hat{\gamma}_{\mathrm{phys}}$ before evaluating the eigenvalue residual
\STATE $\mathcal{L} \leftarrow \lambda_d \|\hat{\gamma} - \gamma_{\mathrm{MEQ}}\|^2 + \lambda_s \mathcal{L}_{\text{spatial}} + \lambda_e \|\mathbf{A}\hat{\mathbf{v}} - \hat{\gamma}_{\mathrm{phys}}\hat{\mathbf{v}}\|^2 / \|\hat{\mathbf{v}}\|^2$
\STATE $\phantom{\mathcal{L}} + \lambda_b \|\widehat{\delta\psi}|_{\text{wall}}\|^2 + \lambda_n (\|\hat{\mathbf{v}}\| - 1)^2 + \lambda_r \|\boldsymbol{\theta}\|^2$
\STATE $\mathcal{L}_{\Xi} \leftarrow \|\widehat{\boldsymbol{\Xi}}_{I_y}-\boldsymbol{\Xi}_{I_y}\|_F^2$ when training the $\boldsymbol{\Xi}_{I_y}$ surrogate
\STATE Update $\boldsymbol{\theta}_{\text{PAT}}$ on $\mathcal{L}$ and $\boldsymbol{\theta}_{\Xi}$ on $\mathcal{L}_{\Xi}$ or the joint objective, depending on the training stage
\end{algorithmic}
\end{algorithm}

\section{Implementation, Training, and Inference}
\label{sec:training}

The implementation uses PyTorch\cite{paszke2019pytorch}. The architecture is intentionally small enough for CPU inference, while the tokenized attention block preserves access to mesh resolved structure.

\label{sec:hyperparams}

All reported experiments use the same compact architecture. The model stacks $L{=}6$ PAT blocks with hidden dimension $C{=}256$, $M{=}64$ learned mesh tokens, and $H{=}8$ attention heads ($C_h{=}32$ per head). Rotary positional embeddings have dimension $C_p{=}32$. The readout head uses $M_q{=}4$ query tokens. Each FFN expands to $C{=}1024$ with GELU activation.

\label{sec:optimization}

The growth rates span two orders of magnitude ($5$ to $650$~s$^{-1}$ for C-Mod; $10$ to $1200$~s$^{-1}$ for SPARC), so all targets are standardized to zero mean and unit variance. Reported dimensional errors are obtained by inverting this transformation.

Training runs for 200 epochs with batch size 32. AdamW\cite{adam_optimizer} is used with cosine learning rate decay\cite{loshchilov2016sgdr} from $10^{-3}$ to $10^{-4}$ after a 5\% warmup, weight decay $8{\times}10^{-4}$, and gradient clipping at norm 1.0. Loss weights $\lambda_*$ are initialized proportional to $1/\mathcal{L}_*^{(0)}$; $\lambda_e$ and $\lambda_b$ are annealed from zero over the first 20 epochs. No dropout is used. Mixed precision FP16 training on a single A100 GPU takes 32~h.

\begin{figure}[htbp]
    \centering
    \subfloat[SPARC dataset\label{fig:training_sparc}]{\includegraphics[width=0.9\linewidth]{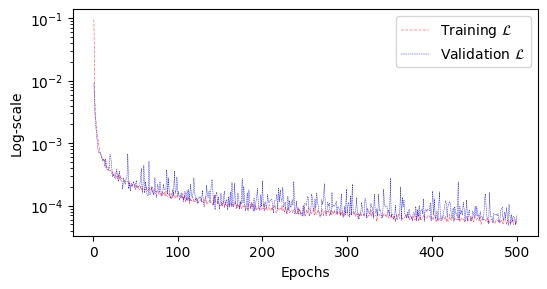}} \\
    \subfloat[C-Mod dataset\label{fig:training_cmod}]{\includegraphics[width=0.9\linewidth]{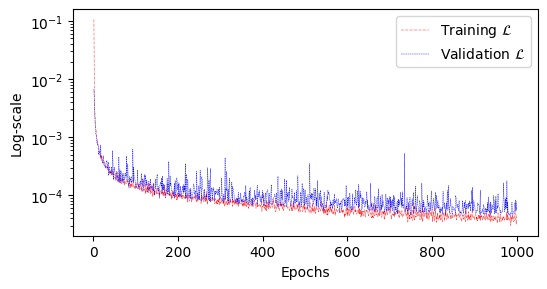}}
    \caption{Training (solid) and validation (dashed) MSE for (a) SPARC and (b) C-Mod. Both converge to ${\sim}10^{-4}$ without divergence.}
    \label{fig:training}
\end{figure}

Both datasets converge to normalized MSE~${\sim}10^{-4}$ in Figure~\ref{fig:training}. SPARC settles at a higher validation loss, consistent with its broader synthetic parameter range and larger extrapolation burden.

\label{sec:ablation_tokens}

The token count study probes the tradeoff between compression and spatial specificity. Accuracy improves from $M{=}1$ (global pooling, $L^2{=}0.015$) to $M{=}256$ ($L^2{=}0.005$), then degrades at $M{=}512$ to $1024$ as assignments fragment into small regions. We use $M{=}64$ ($L^2{=}0.006$), which captures 89\% of the error reduction between $M{=}1$ and $M{=}256$ while using 79\% of the memory required by the $M{=}256$ setting. At the same token count, replacing learned soft assignment with fixed regular grid partitioning increases $L^2$ from 0.006 to 0.011. This comparison supports adaptive tokenization as a useful compression strategy, but it does not isolate every architectural ingredient. Component ablations that remove matrix tokens, the $\widehat{\boldsymbol{\Xi}}_{I_y}$ pathway, the spatial head, or the eigenvalue and boundary losses remain necessary to assign credit to individual mechanisms.

\label{sec:computational_performance}

Latency determines whether PAT is plausible for control cycle use. Measurements use 1{,}000 inference calls after a 100-call warmup with gradients disabled, and Table~\ref{tab:inference} compares PAT with MEQ-FGE-L and RZIP across CPU and GPU settings.

\begin{table*}[htbp]
    \centering
    \caption{Inference latency benchmarks across hardware platforms and comparison with physics based codes. Statistics computed over 1{,}000 inference calls following 100-call warmup.}
    \label{tab:inference}
    \begin{tabular}{lccccc}
        \toprule
        Method & Platform & Mean (ms) & Std (ms) & 99th \%ile (ms) & Hz \\
        \midrule
        MEQ-FGE-L & 1 CPU core & $\sim$180{,}000 & N/A & N/A & 0.006 \\
        RZIP & 1 CPU core & 1.2 & 0.1 & 1.8 & 833 \\
        \midrule
        \textbf{PAT (ours)} & \textbf{1 CPU core} & \textbf{35.2} & \textbf{2.1} & \textbf{41.3} & \textbf{28.4} \\
        PAT & 8 CPU cores & 8.5 & 0.9 & 11.2 & 118 \\
        PAT & V100 (batch=1) & 1.18 & 0.19 & 1.84 & 847 \\
        PAT & V100 (batch=32) & 0.38 & 0.05 & 0.51 & 2{,}632 \\
        \bottomrule
    \end{tabular}
\end{table*}

On one Intel Xeon E5-2680 core, PAT evaluates in 35~ms, a $5{,}100\times$ speedup over MEQ-FGE-L. Eight CPU cores reduce the latency to 8.5~ms, while a V100 GPU reaches 1.18~ms for batch size 1. These measurements make ensemble uncertainty or fast parameter scans plausible, but they do not include every PCS integration cost.

Mesh level equilibrium features, global scalars, and coil currents are produced by real time EFIT (${\sim}1$ to 2~ms). The matrices $\mathbf{M}_{ee}$, $\mathbf{M}_{ey}$, and $\mathbf{R}_e$ are machine constants, precomputed once. Because the $\boldsymbol{\Xi}_{I_y}$ surrogate replaces the ${\sim}3$ min Grad-Shafranov solve with a ${\sim}1$ ms MLP evaluation, the single model chained pipeline in Eq.~\ref{eq:pipeline} runs in under 40~ms on one core. A five member ensemble requires parallel execution or roughly proportional additional latency.

%%%%%%%%%%%%%%%%%%%%%%%%%%%%%%%%%%%%%%%%%%%%%%%%%%%%%%%%%%%%%
\section{Growth Rate Prediction}
\label{sec:growth_rate_performance}

In this section, the scalar $n=0$ axisymmetric growth rate fidelity is evaluated on held out C-Mod and SPARC equilibria, then against a scalar input MLP baseline. The through line is solver fidelity: every metric in this section compares PAT with MEQ-FGE-L labels rather than with measured nonlinear VDE dynamics.

\label{sec:data_splitting}

All scalar prediction metrics use the discharge- or trajectory level split defined above: no C-Mod discharge and no SPARC scenario trajectory contributes slices to more than one partition. Within the machine and scenario strata, the split is ordered chronologically to reduce direct temporal leakage from adjacent time slices. The C-Mod test set contains 1{,}524 equilibria from 47 discharges; the SPARC SSL test set contains 1{,}310 equilibria from 63 scenario trajectories. A Spearman check between test set prediction error and time gap from the training cutoff gives $\rho = 0.03$ ($p = 0.31$), indicating no monotonic performance drift with temporal distance. It does not prove statistical independence, since slices within a discharge remain correlated.

\label{sec:cmod_accuracy}

\begin{figure*}[htbp]
    \centering
    \subfloat[Parity plot: C-Mod test set\label{fig:parity_cmod_sub}]{\includegraphics[width=0.48\linewidth]{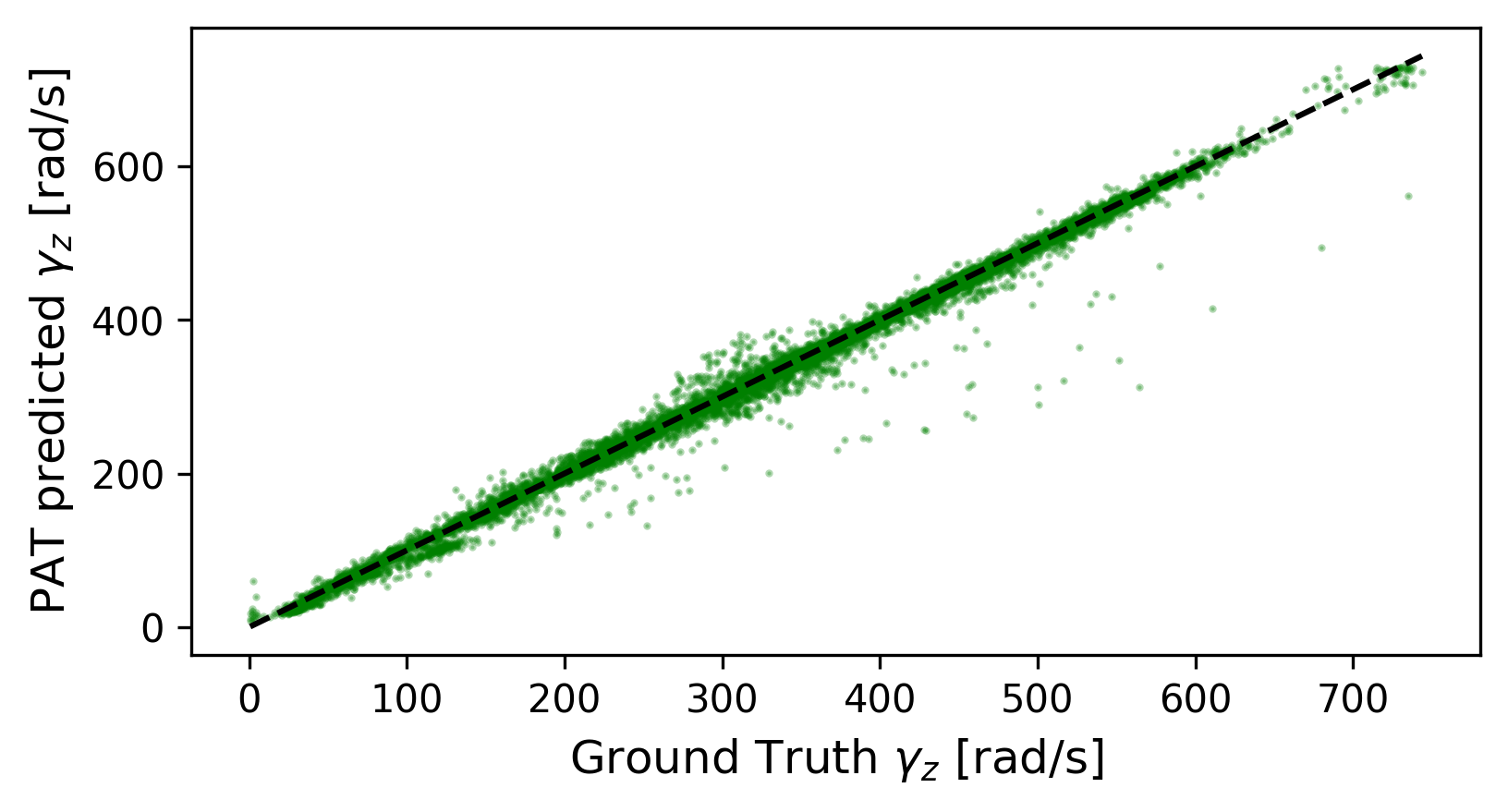}}
    \subfloat[Parity plot: SPARC SSL test set\label{fig:parity_sparc_sub}]{\includegraphics[width=0.48\linewidth]{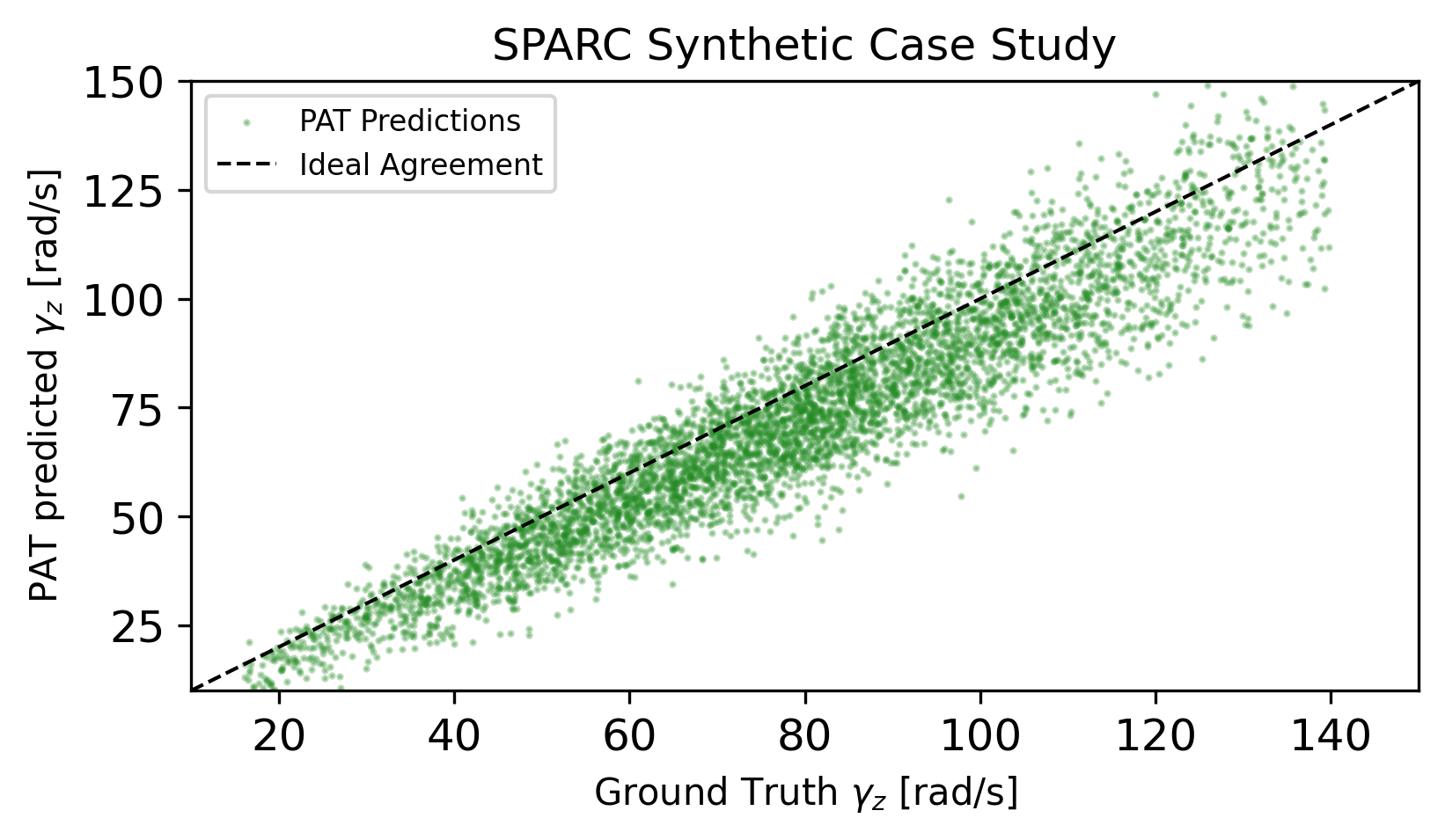}} \\
    \subfloat[Error distribution: C-Mod\label{fig:error_cmod_sub}]{\includegraphics[width=0.48\linewidth]{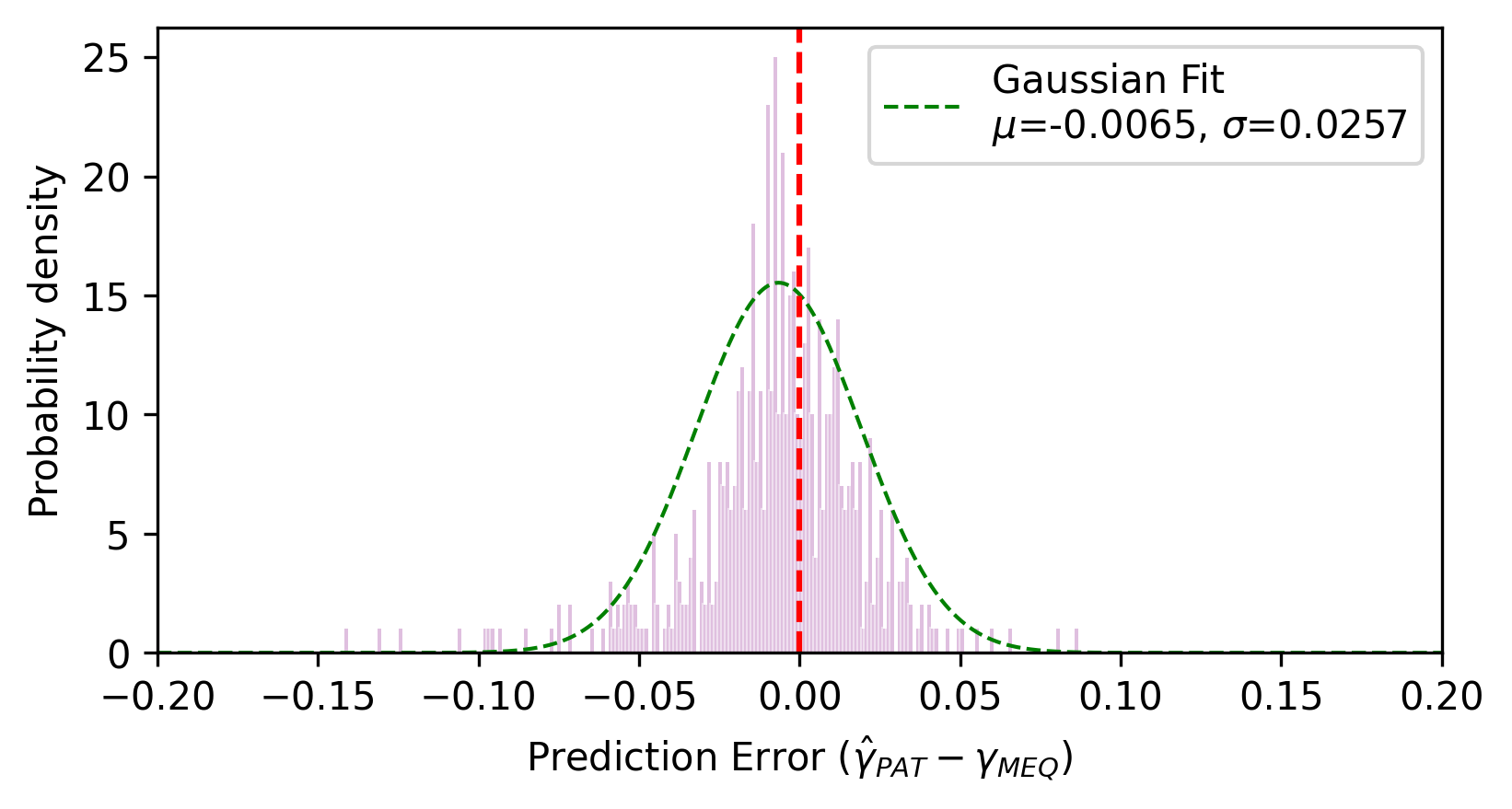}}
    \subfloat[Error distribution: SPARC\label{fig:error_sparc_sub}]{\includegraphics[width=0.48\linewidth]{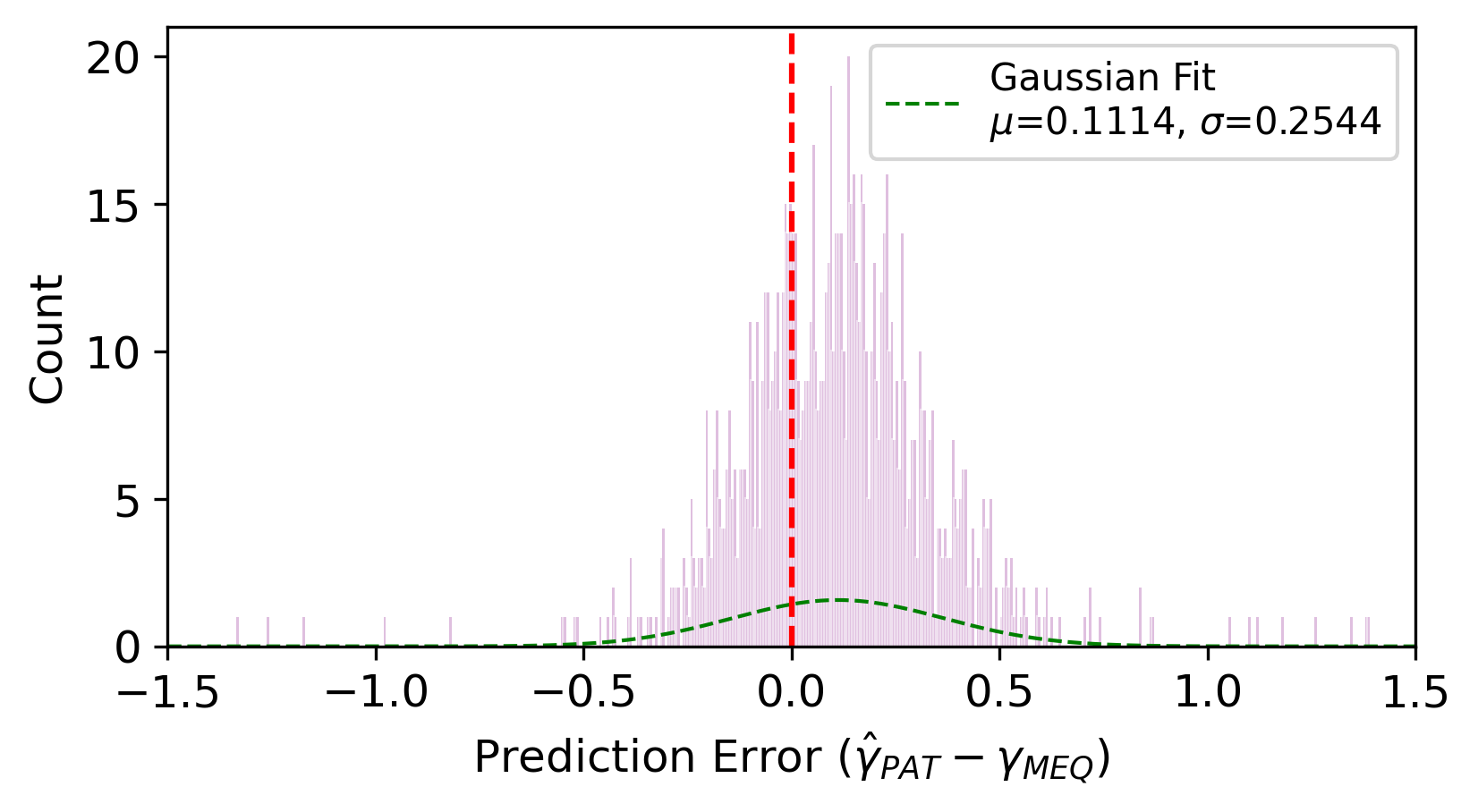}} \\
    \subfloat[PAT vs.\ MLP error comparison\label{fig:pat_vs_mlp}]{\includegraphics[width=0.48\linewidth]{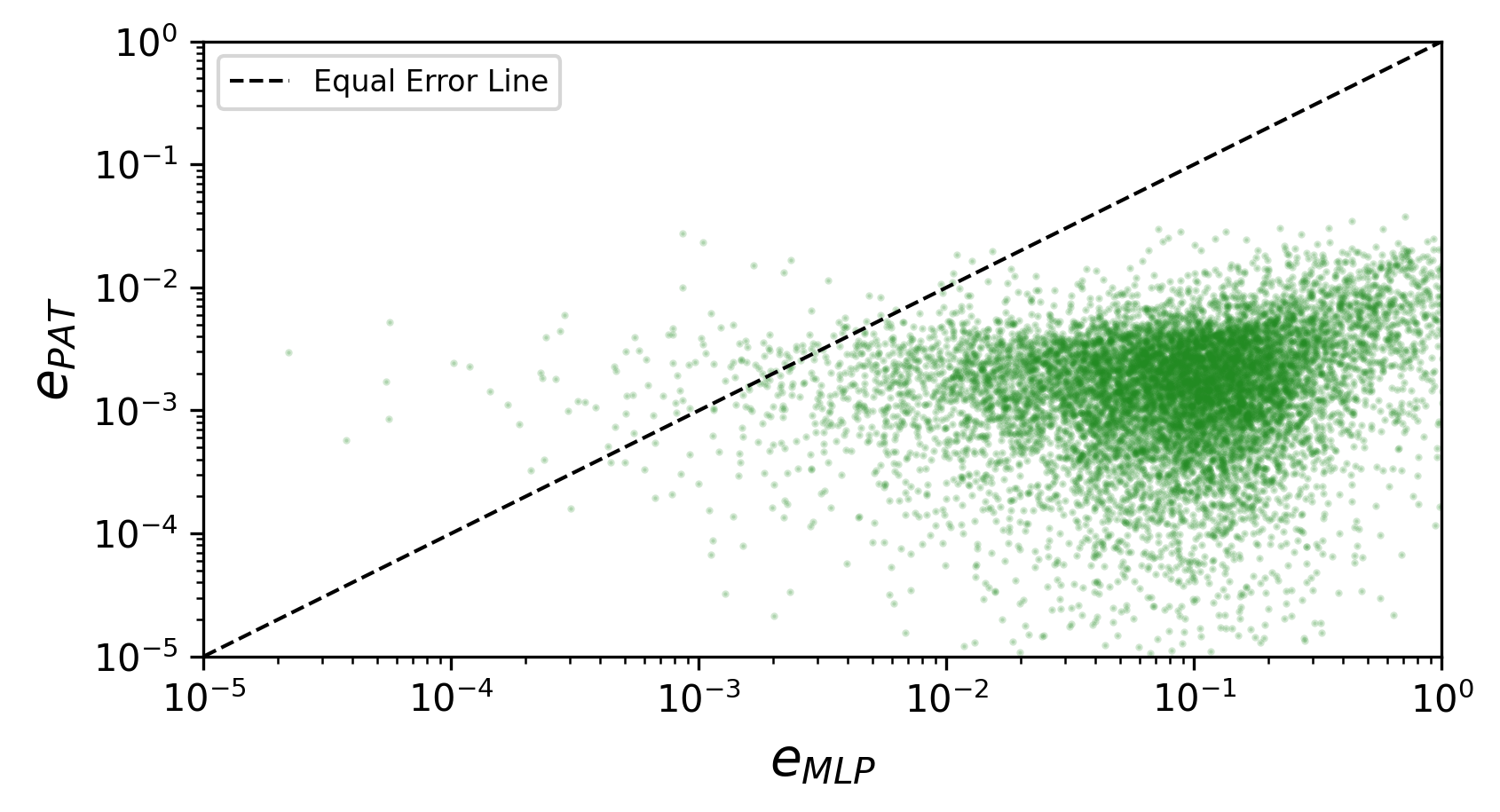}}
    \caption{Growth rate prediction against MEQ-FGE-L labels. (a,b) C-Mod and SPARC parity plots show tight interpolation on C-Mod and broader residuals on the synthetic SPARC domain. (c,d) Residual distributions reflect the same shift. (e) PAT has lower absolute error than the scalar MLP on 94.2\% of test equilibria, with the largest gains in cases where mesh resolved structure matters.}
    \label{fig:growth_rate_summary}
\end{figure*}

C-Mod is the closest interpolation test within the MEQ-FGE-L labeled experimental equilibrium distribution, because the held out cases remain inside the operating range represented in training. Across Ohmic through H-mode cases with $\kappa \in [1.50, 1.82]$ and $I_p \in [0.4, 1.8]$~MA, the parity plot in Figure~\ref{fig:growth_rate_summary}(a) gives $R^2 = 0.997$, with agreement extending from marginal cases ($\gamma_{\mathrm{MEQ}} < 20$~s$^{-1}$) to strongly unstable equilibria ($\gamma_{\mathrm{MEQ}} > 600$~s$^{-1}$).

The residual histogram in Figure~\ref{fig:growth_rate_summary}(c) is nearly Gaussian, with $\mu_\epsilon = -0.007$ and $\sigma_\epsilon = 0.026$ in normalized units. After inverting the target standardization, the MAE is 5.4~s$^{-1}$. Stratification shows little variation across elongation ($\Delta$MAE~$< 1.2$~s$^{-1}$) or current ($<0.9$~s$^{-1}$). H-mode equilibria are somewhat harder than Ohmic equilibria, with MAE increasing from 4.2 to 6.8~s$^{-1}$.

\label{sec:sparc_accuracy}

The SPARC test set is harder because the synthetic library reaches higher elongation and broader pressure/current profile space ($\kappa$ up to 2.35, $\beta_N$ up to 2.8). Panels (b) and (d) of Figure~\ref{fig:growth_rate_summary} show the SPARC parity plot and residual distribution. On the full SPARC SSL test set, PAT obtains $R^2 = 0.94$ and MAE~$=12.7$~s$^{-1}$. The largest residuals occur in high elongation double null cases with high triangularity and small plasma wall gap, where passive conductor coupling and the effective inductance $\mathbf{L}_{ee}^*$ are most sensitive to geometry. The same region appears again in the error stratification and uncertainty analysis, so these equilibria are natural candidates for fallback evaluation rather than blind deployment.

The full test SPARC result should be separated from the benchmark overlap transfer test. On the full SPARC SSL test set, the jointly trained model gives the 12.7~s$^{-1}$ MAE reported above. On the smaller overlap subset used later for zero shot transfer,  C-Mod only training gives MAE 18.7~s$^{-1}$, while joint C-Mod/SPARC training reduces that subset error to 8.1~s$^{-1}$.

\label{sec:mlp_comparison}

To isolate the contribution of the mesh aware architecture, we compare PAT with a Multi Layer Perceptron baseline that uses only global scalar parameters (Appendix~\ref{appendixA}) as input. The MLP architecture and training procedure are detailed in Appendix~\ref{appendixA}.

Panel (e) of Figure~\ref{fig:growth_rate_summary} compares PAT and MLP errors point by point. Points below the diagonal have lower PAT error than MLP error; 94.2\% of test equilibria fall in this region. The largest gains occur for cases with spatially complex current distributions, where scalar shape parameters are a narrow description of the actual electromagnetic state.

\label{sec:bootstrap}

To assess sensitivity to the sampled test equilibria, we computed 95\% bootstrap confidence intervals using $B = 1{,}000$ equilibrium level resampling iterations with replacement from each test set. Table~\ref{tab:bootstrap_ci} reports the resulting intervals for $R^2$, MAE, RMSE, and maximum absolute error. Because this bootstrap is not clustered by discharge or synthetic trajectory, the CIs should be read as conditional on the sampled equilibria and not as a full correction for intra shot autocorrelation.

\begin{table*}[htbp]
\centering
\caption{Performance metrics with 95\% bootstrap confidence intervals ($B = 1{,}000$ samples). Intervals estimated as 2.5th and 97.5th percentiles of the bootstrap distribution.}
\label{tab:bootstrap_ci}
\begin{tabular}{lccc}
\toprule
Metric & C-Mod Test & SPARC SSL Test & Combined \\
\midrule
$R^2$ & 0.997 [0.995, 0.998] & 0.940 [0.920, 0.960] & 0.976 [0.968, 0.982] \\
MAE (s$^{-1}$) & 5.4 [4.8, 6.1] & 12.7 [10.9, 14.6] & 8.7 [7.9, 9.6] \\
RMSE (s$^{-1}$) & 8.3 [7.4, 9.3] & 21.4 [18.6, 24.3] & 15.2 [13.8, 16.8] \\
Max $|\epsilon|$ (s$^{-1}$) & 42 [38, 48] & 87 [76, 102] & 87 [79, 98] \\
\bottomrule
\end{tabular}
\end{table*}

\begin{figure}[htbp]
\centering
\includegraphics[width=0.48\textwidth]{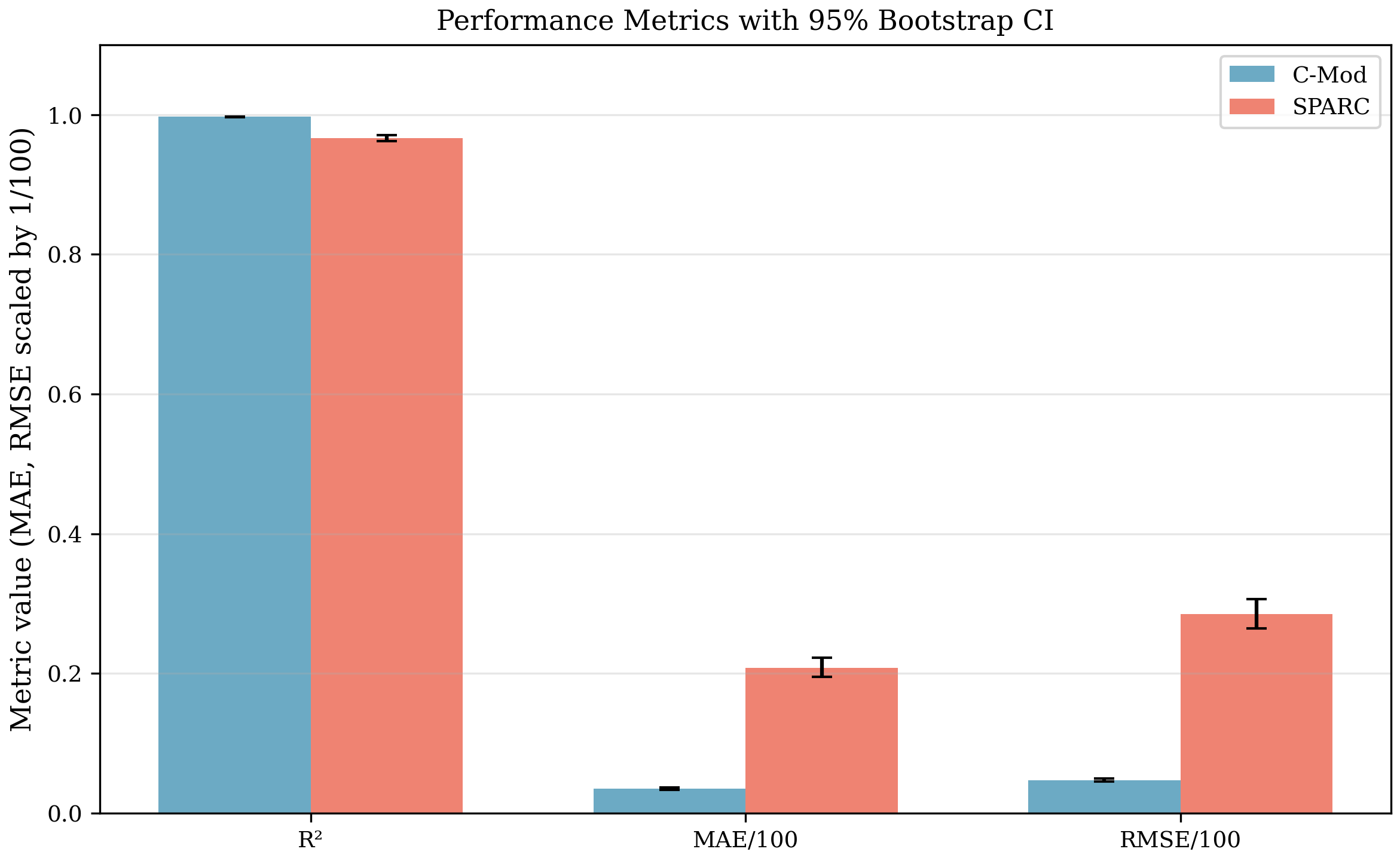}
\caption{Bootstrap confidence intervals for the main growth rate metrics. Wider SPARC intervals reflect the broader synthetic operating space.}
\label{fig:bootstrap_ci}
\end{figure}

The tight C-Mod intervals ($R^2 \in [0.995, 0.998]$) suggest stable solver fidelity within the sampled held out equilibria. SPARC intervals are wider, reflecting both smaller sample size and greater variation in the synthetic extrapolation regime. Figure~\ref{fig:bootstrap_ci} visualizes the same confidence intervals.

We also tested the null hypothesis $H_0$: PAT and MLP have equal MSE against the one sided alternative $H_1$: PAT has lower MSE. The equilibrium level bootstrap statistic $t = (\text{MSE}_{\text{MLP}}- \text{MSE}_{\text{PAT}}) / \text{SE}_{\text{bootstrap}} = 12.8$ yields $p < 0.001$ (one tailed), providing evidence that the accuracy difference is unlikely to be attributable to equilibrium level resampling variability alone. As with the CIs above, this test is not clustered by discharge or trajectory.

A Sobol sensitivity analysis~\cite{sobol2001global} ($N_s = 50{,}000$ samples) ranks elongation first ($S_\kappa = 0.34$), followed by internal inductance ($S_{\ell_i} = 0.21$) and plasma wall gap ($S_{L_{\text{gap}}} = 0.18$). These rankings agree with the integrated gradient analysis later in the manuscript and are consistent with known vertical stability drivers. They do not rule out correlated proxies or dataset specific structure.

\section{Spatial Eigenfunction Reconstruction}
\label{sec:spatial_reconstruction}

A scalar growth rate indicates how fast an instability grows; control design also needs the associated mode structure. We therefore evaluate the decoded $\delta j_\phi(R,Z)$ and the derived perturbed flux $\delta\psi_p$ on static equilibria and on a time evolving VDE.

\label{sec:static_eigenfunction}

\begin{figure*}[htbp]
    \centering
    \subfloat[MEQ-FGE-L label\label{fig:jpar_meq}]{%
        \includegraphics[width=0.7\linewidth]{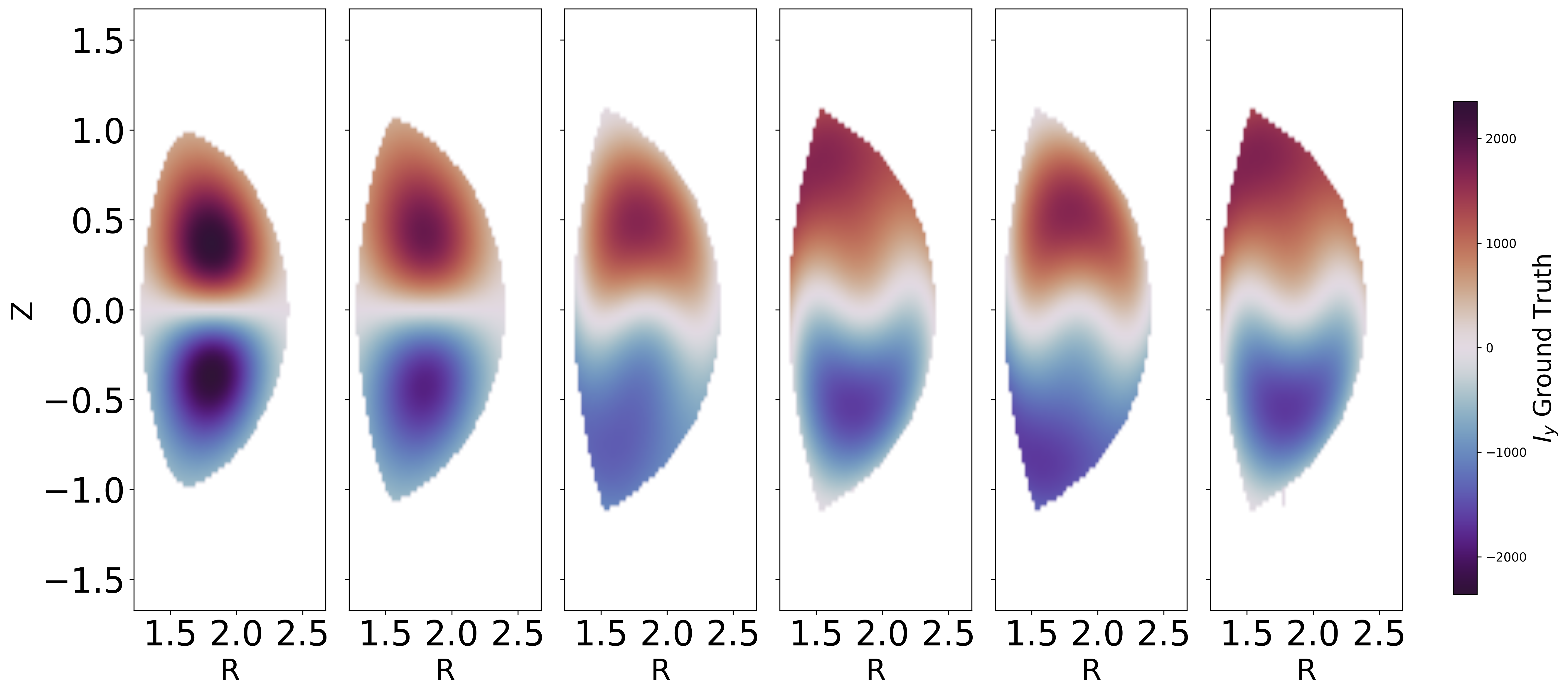}}\\
    \subfloat[PAT prediction\label{fig:jpar_network}]{%
        \includegraphics[width=0.71\linewidth]{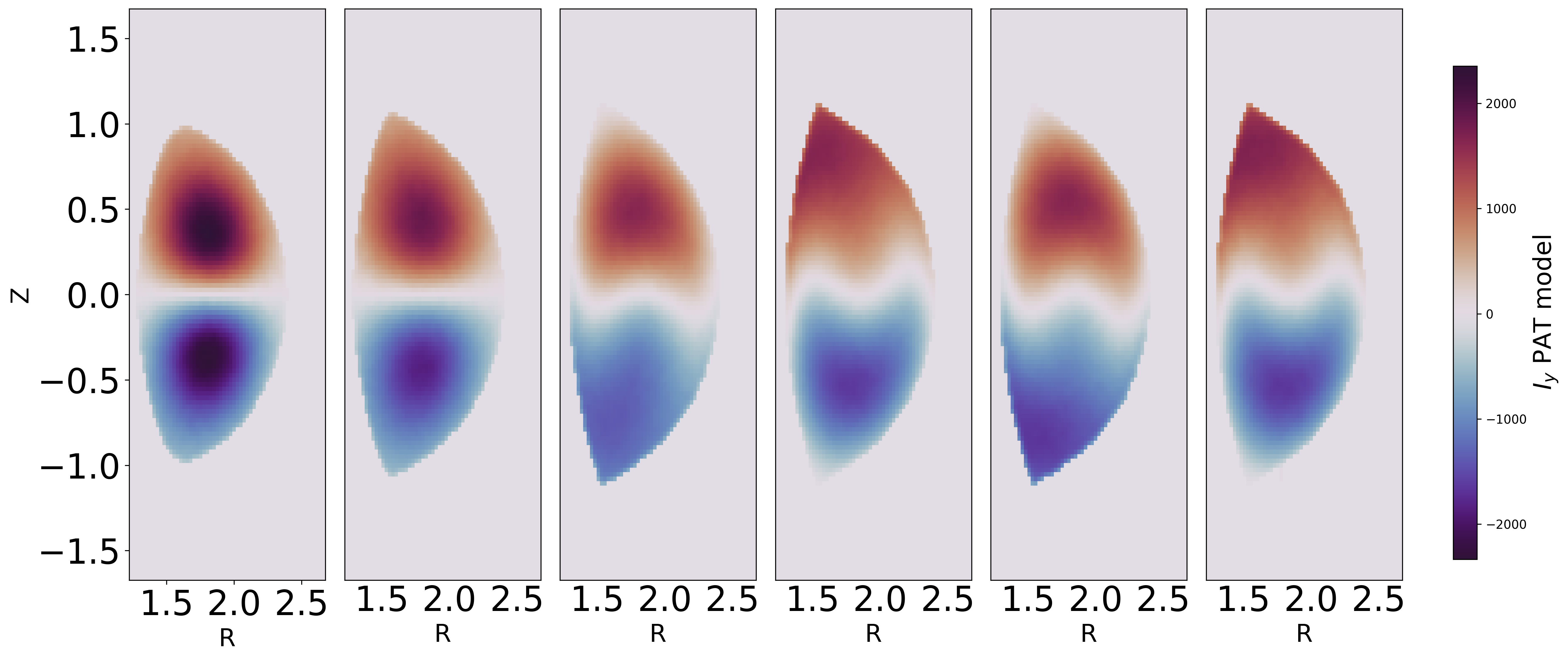}}\\
    \subfloat[Residual error\label{fig:jpar_error}]{%
        \includegraphics[width=0.71\linewidth]{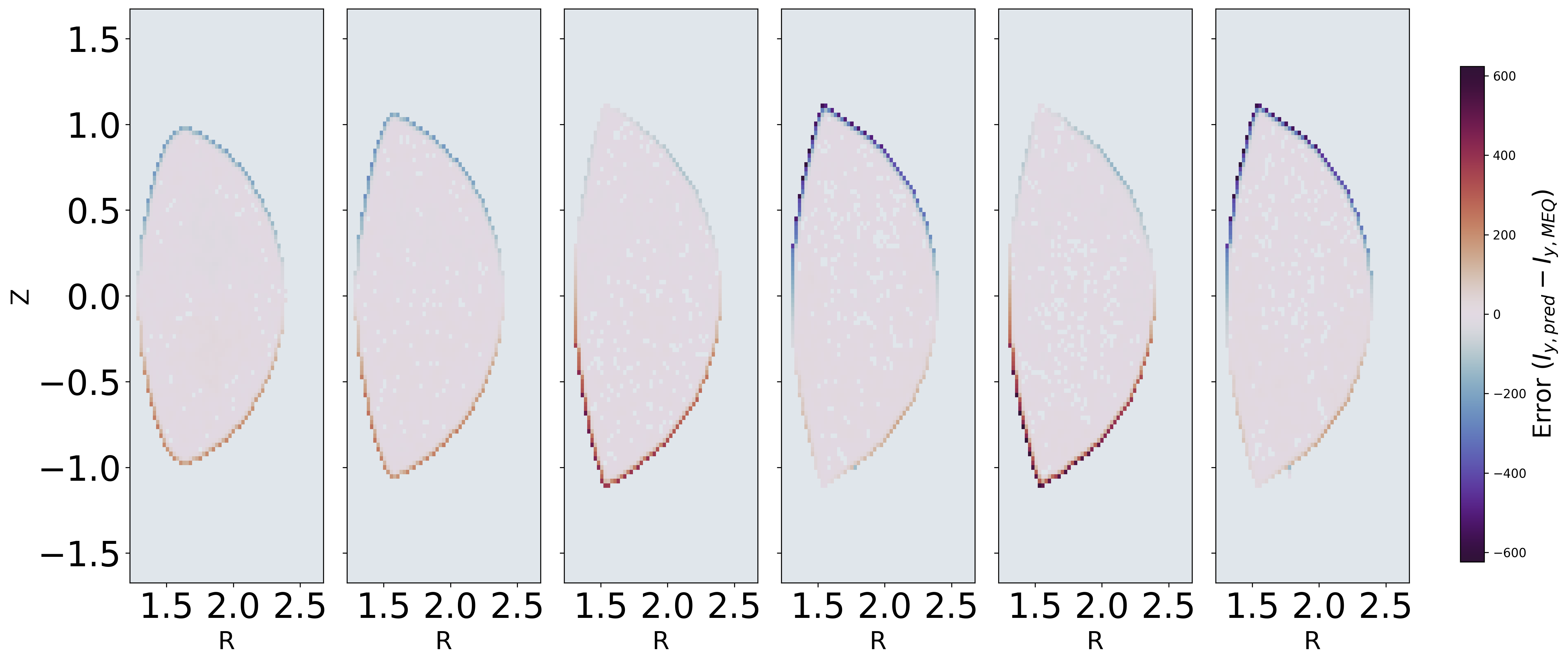}}
    \caption{$\delta j_\phi(R,Z)$ for six SPARC double null equilibria. (a) MEQ-FGE-L. (b) PAT. (c) Signed residual; errors concentrate near the separatrix and steep gradient regions.}
    \label{fig:jpar_comparison}
\end{figure*}

Static reconstruction is evaluated on six SPARC double null equilibria with $\kappa \in [1.75, 2.35]$ and growth rates from 45 to 480~s$^{-1}$. The object being reconstructed is the normalized dominant linear eigenfunction from MEQ-FGE-L, not a measured nonlinear VDE current redistribution. Figure~\ref{fig:jpar_comparison} compares PAT predicted $\delta j_\phi(R,Z)$ against those solver fields.

The comparison checks three aspects of the solver computed mode. First, PAT preserves the odd vertical parity structure of the $n=0$ perturbation after applying the sign convention described in Eq.~\ref{eq:eigenfunction}; the absolute positive/negative orientation is not itself physical because a linear eigenfunction is defined only up to sign. Second, the normalized amplitude and spatial extent are close in the examples shown, with peak current density perturbations matched within roughly 10\%. This supports reconstruction of the MEQ-FGE-L eigenfunction under the chosen normalization, but it does not independently set physical eddy current amplitudes. Third, the X point and divertor region structure in double null configurations is reproduced, indicating that the learned representation preserves topology correlated spatial features present in the inputs without proving magnetic connectivity learning.

The residual map in Figure~\ref{fig:jpar_comparison}(c) concentrates near the separatrix and mesh regions adjacent to conductor boundaries, where spatial gradients are steepest. The largest absolute errors reach $\sim 200$~A/m$^2$ ($\sim 10\%$ of local field strength). One possible contributor is smoothing from the soft assignment tokenization in Eq.~\ref{eq:soft_assignment}; basis interpolation, mesh resolution, and mismatch between the MEQ-FGE-L and PAT sampling grids may also contribute.

Quantitative metrics across both test sets are summarized in Table~\ref{tab:spatial_metrics_summary}:

\begin{table*}[htbp]
\centering
\caption{Spatial eigenfunction reconstruction metrics. Statistics reported as median [95\% bootstrap confidence interval].}
\label{tab:spatial_metrics_summary}
\begin{tabular}{lccc}
\toprule
Metric & C-Mod Test & SPARC SSL Test & Combined \\
\midrule
$\epsilon_{L^2}(\delta j_\phi)$ & 0.049 [0.046, 0.052] & 0.059 [0.054, 0.065] & 0.053 [0.050, 0.057] \\
Dice$(\delta j_\phi)$ & 0.92 [0.90, 0.93] & 0.87 [0.84, 0.90] & 0.90 [0.88, 0.92] \\
SSIM$(\delta j_\phi)$ & 0.94 [0.93, 0.95] & 0.91 [0.89, 0.93] & 0.93 [0.91, 0.94] \\
\midrule
$\epsilon_{L^2}(\delta\psi_p)$ & 0.042 [0.039, 0.045] & 0.054 [0.049, 0.060] & 0.047 [0.044, 0.051] \\
SSIM$(\delta\psi_p)$ & 0.95 [0.94, 0.96] & 0.93 [0.91, 0.94] & 0.94 [0.93, 0.95] \\
\bottomrule
\end{tabular}
\end{table*}

\begin{figure*}[htbp]
    \centering
    \subfloat[MEQ-FGE-L $\delta\psi_p$\label{fig:flux_meq}]{%
        \includegraphics[width=0.6\textwidth]{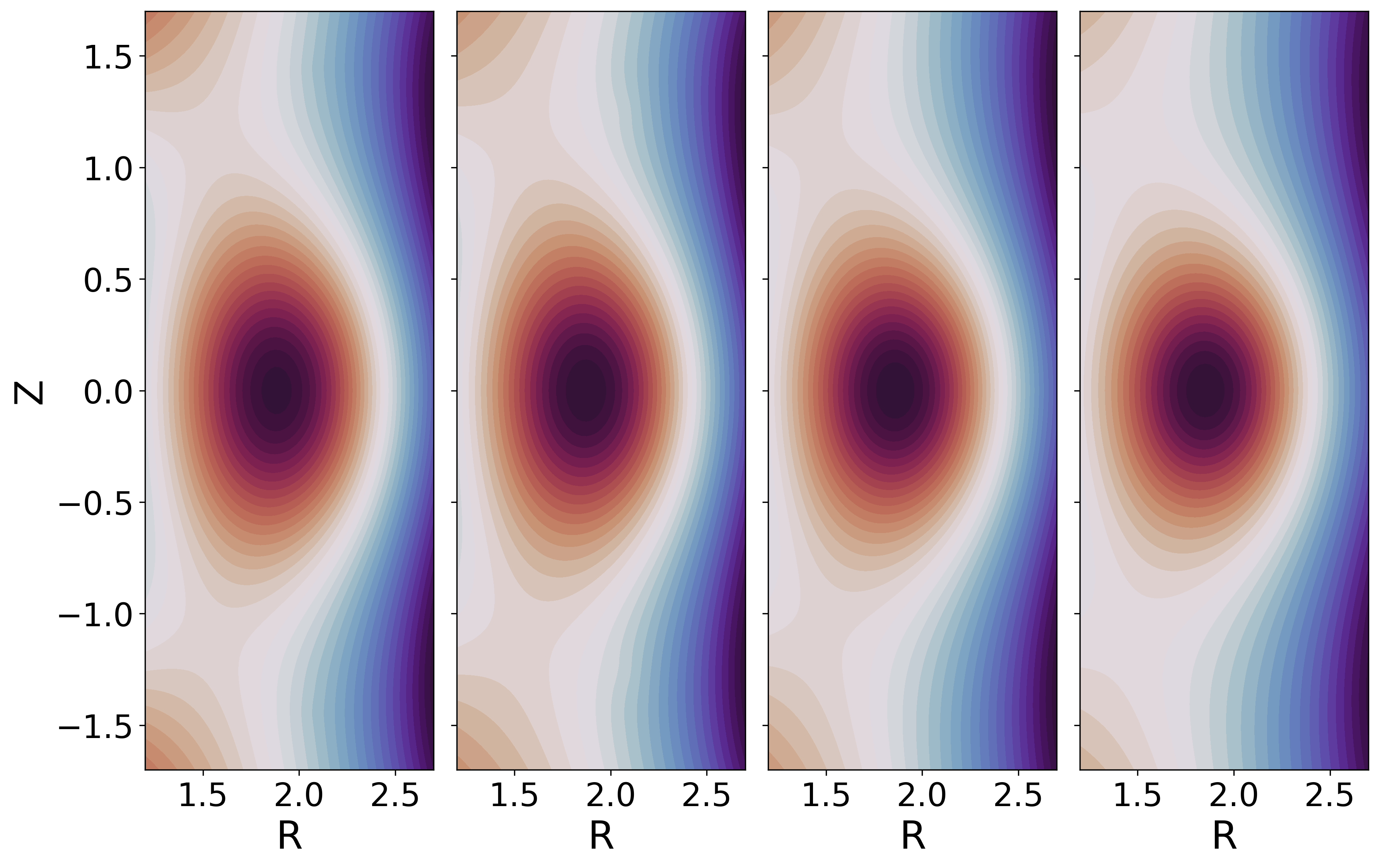}}\\
    \subfloat[PAT $\delta\psi_p$\label{fig:flux_pat}]{%
        \includegraphics[width=0.6\textwidth]{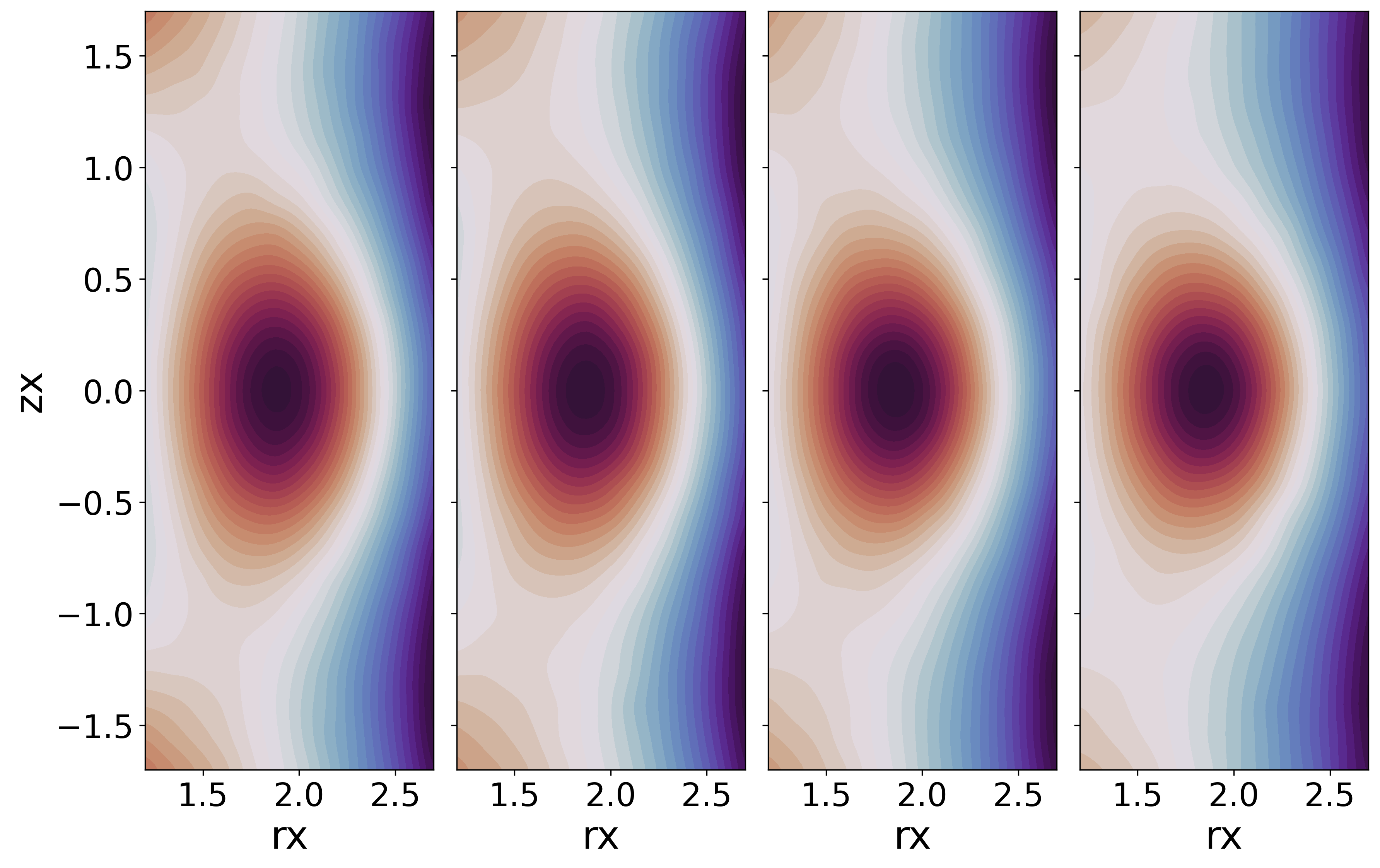}}\\
    \subfloat[Absolute error\label{fig:flux_error}]{%
        \includegraphics[width=0.6\textwidth]{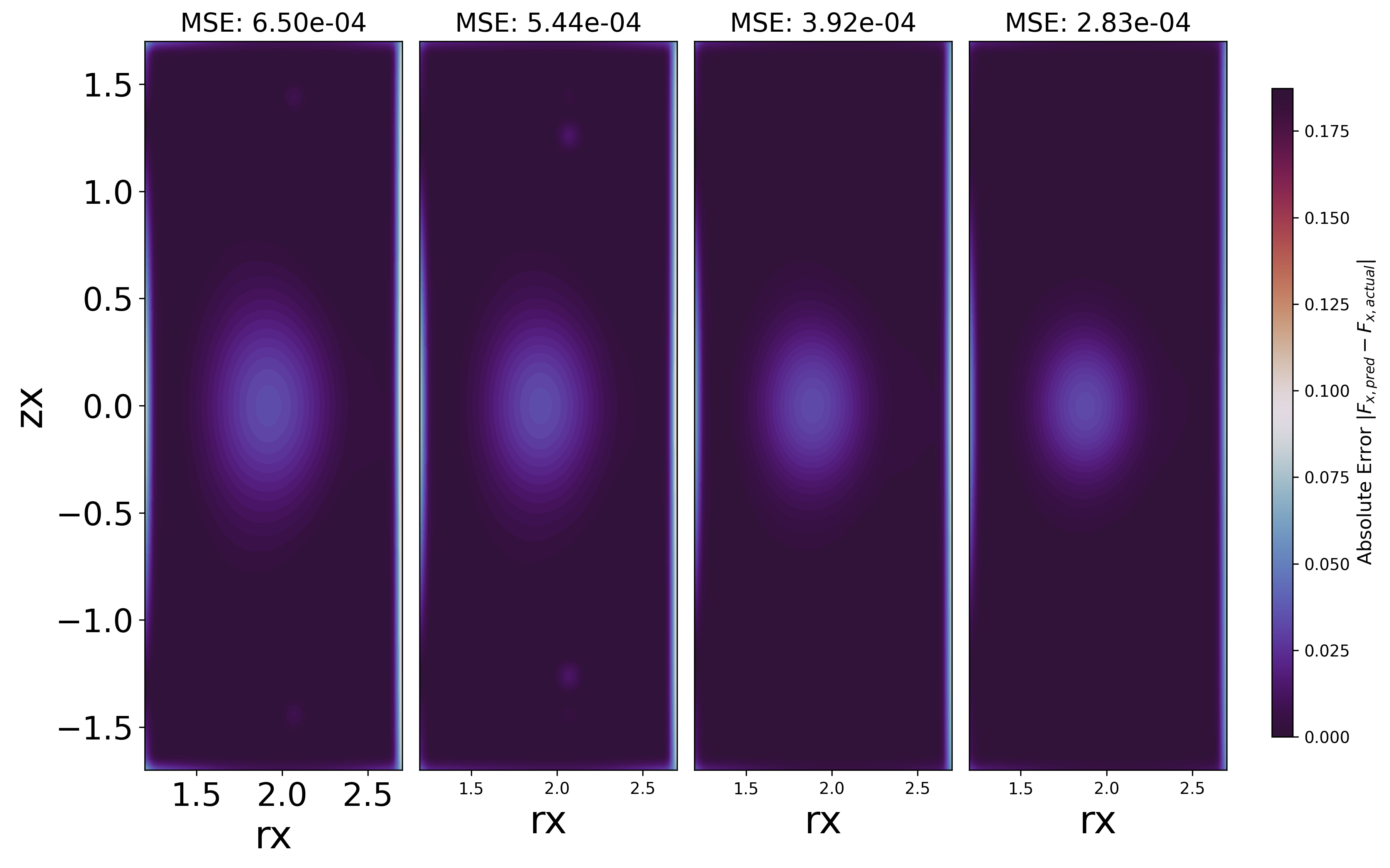}}
    \caption{Perturbed poloidal flux reconstruction for representative SPARC equilibria. The comparison checks the derived flux field used in boundary loss evaluation and complements the $\delta j_\phi$ comparison in Figure~\ref{fig:jpar_comparison}.}
    \label{fig:flux_comparison}
\end{figure*}

Performance degrades modestly on the SPARC SSL test set compared with C-Mod within distribution predictions, with $\Delta\epsilon_{L^2} \approx 0.01$ across both field quantities. This is still agreement with synthetic MEQ-FGE-L labels, not validation of SPARC hardware wall physics. Figure~\ref{fig:flux_comparison} gives the corresponding visual check for $\delta\psi_p$, the derived flux field used when evaluating the boundary loss term.

\label{sec:hot_vde_case}

The hot VDE case then asks whether repeated static PAT evaluations remain coherent during a transient. We applied the model offline to C-Mod shot \#1120427020, which belongs to the held out test partition and was not seen during training. The plasma descends by roughly 45~cm over 450~ms before striking the lower divertor, corresponding to an average vertical speed of about 1~m/s. Each time slice is treated as a reconstructed equilibrium for comparison with MEQ-FGE-L. That snapshot view is locally useful before topology change and halo current physics dominate.

\begin{figure*}[htbp]
    \centering
    \subfloat[MEQ-FGE-L label]{%
        \includegraphics[width=0.7\linewidth]{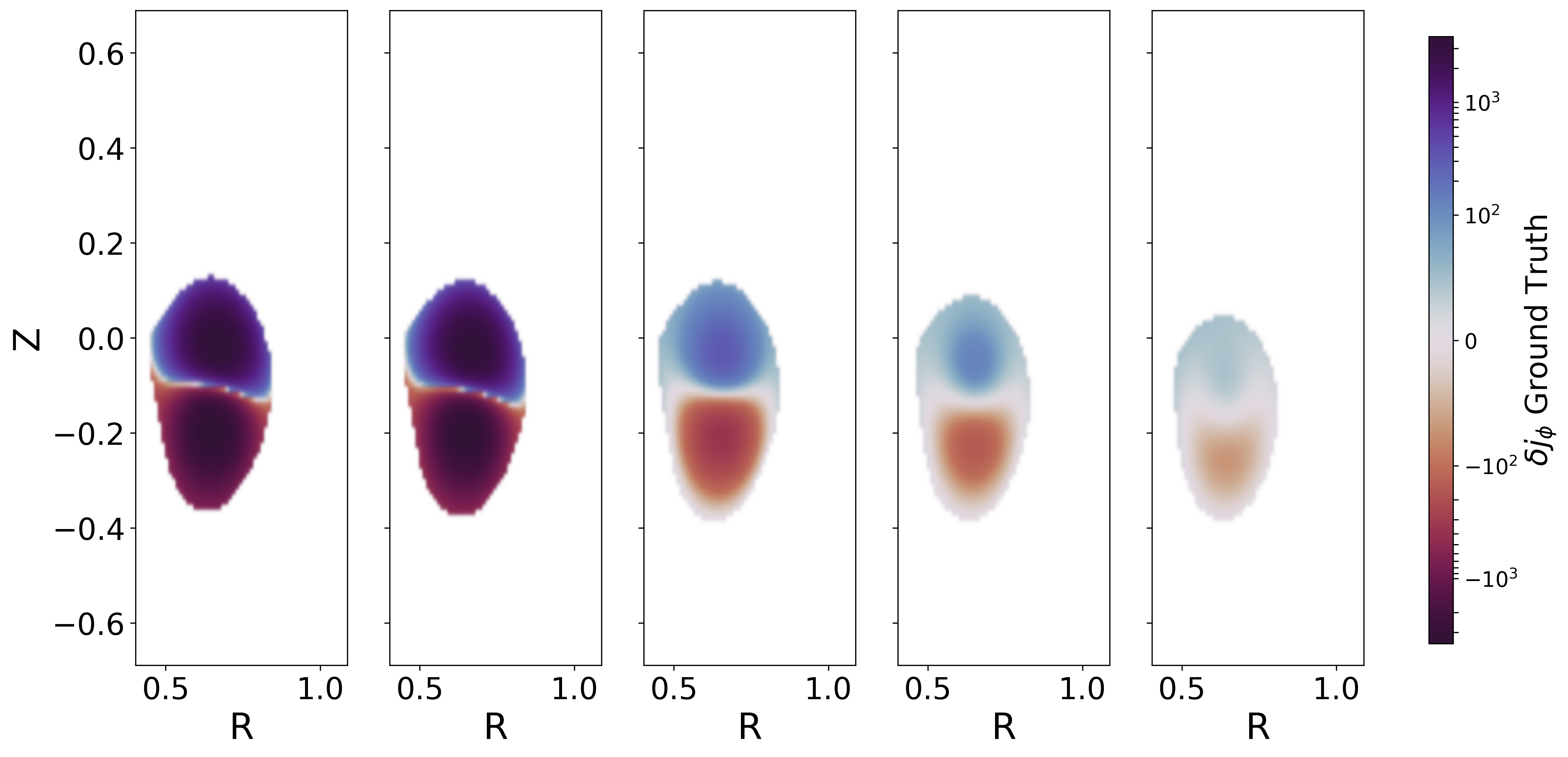}}\\
    \subfloat[PAT prediction]{%
        \includegraphics[width=0.7\linewidth]{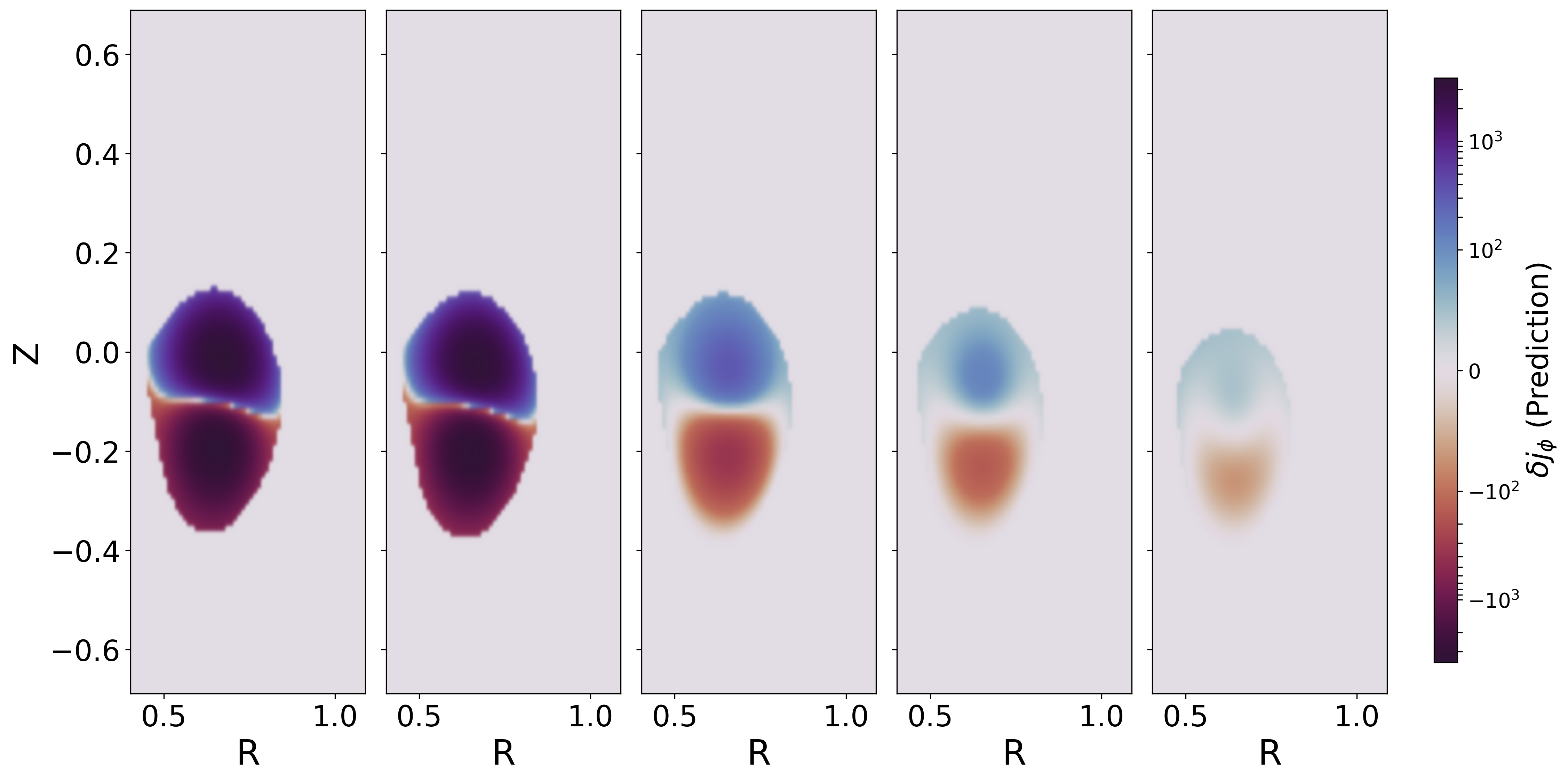}}\\
    \subfloat[Absolute error]{%
        \includegraphics[width=0.7\linewidth]{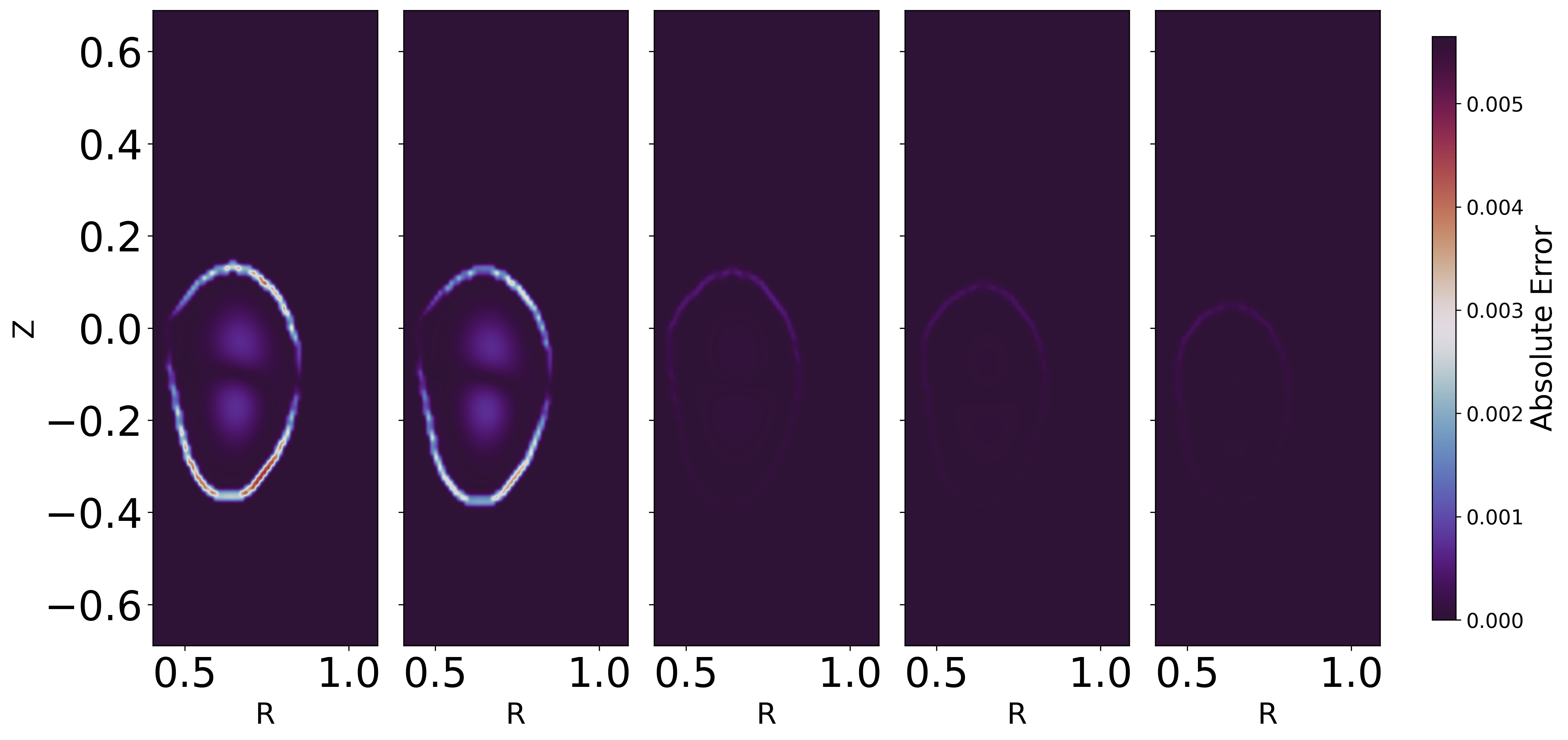}}
    \caption{Hot VDE time evolution (shot \#1120427020, test partition). Top: MEQ-FGE-L. Middle: PAT. Bottom: absolute error. Columns at 100~ms intervals. Errors decrease from MSE $6.5{\times}10^{-4}$ to $2.8{\times}10^{-4}$.}
    \label{fig:vde_time_evolution}
\end{figure*}

The field snapshots in Figure~\ref{fig:vde_time_evolution} show that the bipolar perturbation follows the plasma's downward motion: the peak positive perturbation moves from $Z \approx -0.1$~cm at $t=0$ to $Z \approx -0.19$~cm by $t=400$~ms. The growth rate increases from $\hat{\gamma} = 120$~s$^{-1}$ to $\hat{\gamma} = 385$~s$^{-1}$ as the lower gap shrinks and the vertical eigenmode couples more strongly to lower passive structures; PAT tracks this threefold solver label increase with MAE~$<8$~s$^{-1}$ at all time points. Prediction errors decrease over the sequence (MSE from $6.5 \times 10^{-4}$ at $t=0$ to $2.8 \times 10^{-4}$ at $t=400$~ms). One possible explanation is a stronger weighting of lower passive structures late in the event, but training set density and reconstruction quality may also vary with vertical position.
\par
Against MEQ-FGE-L, the surrogate maintains $R^2 > 0.98$ over the 450~ms window. This is a test of temporal consistency under repeated static evaluations, not a validation against nonlinear VDE dynamics.

%%%%%%%%%%%%%%%%%%%%%%%%%%%%%%%%%%%%%%%%%%%%%%%%%%%%%%%%%%%%%
\section{Benchmarking}
\label{sec:benchmarking}
This section present our benchmarking exercises. Our benchmarking separates the accuracy gain from speed and architecture choices by comparing PAT with reduced physics models and neural operator baselines models.

\label{sec:physics_comparison}

First and foremost, the reduced model:  RZIP\cite{Humphreys_2009} treats the plasma as a rigid body, yielding $\gamma_{\text{RZIP}} = (n_s - 1)/(n_s f_s \tau_w)$ where $n_s$ is the passive stability index, $f_s$ is the feedback/stability normalization factor used in the RZIP estimate, and $\tau_w$ is the wall time constant. It evaluates in ${\sim}1.2$~ms but neglects nonrigid current redistribution, which can modify growth rates by factors of 1.5 to 2.0\cite{welander_nonrigid_2005}.

On the direct code comparison subset of $N{=}847$ equilibria, Table~\ref{tab:physics_comparison} compares PAT and RZIP against MEQ-FGE-L; this subset is not the same aggregation as the full C-Mod/SPARC test set in Table~\ref{tab:bootstrap_ci}. RZIP systematically underestimates growth rates for the most unstable equilibria ($\gamma_{\mathrm{MEQ}} > 400$~s$^{-1}$), a regime where high elongation, small wall gaps, and nonrigid response are expected to matter most.

\begin{table}[htbp]
\centering
\caption{Vertical growth rate prediction accuracy against MEQ-FGE-L labels on the direct code comparison benchmark subset ($N = 847$). Speedup is relative to MEQ-FGE-L single core evaluation ($\sim$180~s).}
\label{tab:physics_comparison}
\begin{tabular}{lccccc}
\toprule
Model & MAE & RMSE & $R^2$ & Max $|\epsilon|$ & Speedup \\
& (s$^{-1}$) & (s$^{-1}$) & & (s$^{-1}$) & vs MEQ \\
\midrule
MEQ-FGE-L &   &   & 1.000 & 0 & 1$\times$ \\
RZIP & 31.7 & 47.3 & 0.82 & 187 & 150{,}000$\times$ \\
\textbf{PAT (ours)} & \textbf{5.4} & \textbf{8.3} & \textbf{0.994} & \textbf{42} & \textbf{5{,}100$\times$} \\
\bottomrule
\end{tabular}
\end{table}

The lower error relative to RZIP shows that PAT better reproduces the nonrigid MEQ-FGE-L labels than the rigid body estimate. The present comparison does not isolate attention as the sole cause; the gain is consistent with using mesh resolved equilibrium fields and electromagnetic matrix embeddings that are absent from RZIP.

\begin{figure*}[htbp]
    \centering
    \includegraphics[width=\textwidth]{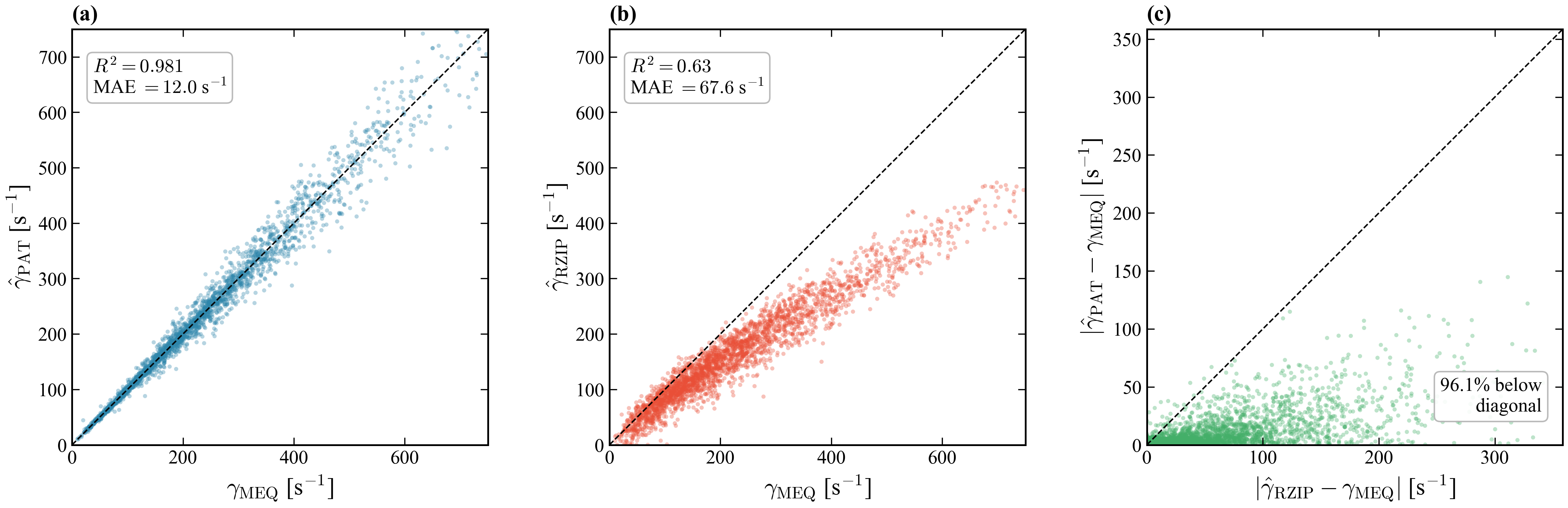}
    \caption{(a) PAT vs.\ MEQ-FGE-L ($R^2 = 0.994$). (b) RZIP vs.\ MEQ-FGE-L ($R^2 = 0.82$). (c) Absolute error comparison on the direct benchmark subset.}
    \label{fig:parity_plots}
\end{figure*}

The same comparison appears as parity plots and direct absolute error comparisons in Figure~\ref{fig:parity_plots}.

For control applications, the dimensionless product $\hat{\gamma}\tau_w$ is a useful indicator of vertical stability margin, although stabilizability also depends on coil authority, sensor latency, voltage and current limits, controller design, and the wall mode spectrum. Small values generally indicate greater routine feedback margin; large values demand faster and more robust active response. SPARC targets $\hat{\gamma}\tau_w \lesssim 3$ during flat top.

We computed $(\hat{\gamma}\tau_w)_{\text{pred}} = \hat{\gamma}_{\text{PAT}} \cdot \tau_w^{\text{MEQ}}$ using the MEQ-FGE-L wall time constant and compared against $\gamma_{\text{MEQ}} \cdot \tau_w^{\text{MEQ}}$. This rescaling preserves the correlation structure of the growth rate prediction ($R^2 = 0.991$) and acts mainly as a consistency check. The modest scatter (RMSE~$= 0.18$ in dimensionless units) comes from growth rate error propagated through the multiplicative rescaling; it does not independently validate the wall time model.

\label{sec:neural_comparison}

\textbf{For the operator models:} we compare PAT against Fourier Neural Operator (FNO)~\cite{Li2020fno, Li2021geo} and DeepONet~\cite{lu2021learning} to test whether learned mesh tokens plus electromagnetic matrix embeddings provide a useful inductive bias relative to spectral or encoder decoder approaches for the vertical stability problem.

\begin{enumerate}
    \item \textbf{FNO-2D}: Four Fourier layers retaining the 12 lowest frequency modes in both $R$ and $Z$ directions, with equilibrium fields lifted to a 64-channel hidden representation. Global pooling followed by MLP produces the scalar growth rate. This architecture has shown competitive performance on plasma evolution modeling~\cite{Gopakumar_2024}.
\item \textbf{DeepONet}: Branch network (convolutional encoder on equilibrium fields) projects to 128-dimensional latent space. Since the output is scalar, the trunk reduces to a learned bias.
\end{enumerate}

Both baselines were trained on identical data splits using the same computational budget (32 GPU hours) as PAT. We note that equal GPU hours does not guarantee equal effective capacity across architectures with different parameter efficiencies; in Table~\ref{tab:neural_operator_comparison}, we report parameter counts to aid interpretation.

\begin{table}[htbp]
\centering
\caption{Neural operator comparison on combined test set. All models trained with identical data splits and computational budget (32 GPU hours). Dashes indicate outputs not produced by the MLP baseline.}
\label{tab:neural_operator_comparison}
\begin{tabular}{lccccc}
\toprule
Architecture & Params & $L^2(\hat{\gamma})$ & Dice$(\delta j_\phi)$ & SSIM & Inference \\
& (M) & & & & (ms) \\
\midrule
FNO-2D & 2.1 & 0.0071 & 0.84 & 0.88 & 42 \\
DeepONet & 1.8 & 0.0083 & 0.81 & 0.85 & 28 \\
MLP (scalars) & 0.3 & 0.0089 & N/A & N/A & 0.5 \\
\textbf{PAT (ours)} & \textbf{2.4} & \textbf{0.0054} & \textbf{0.91} & \textbf{0.93} & \textbf{35} \\
\bottomrule
\end{tabular}
\end{table}

PAT yields the lowest growth rate $L^2$ error (0.0054), 24\% below FNO and 35\% below DeepONet. The advantage also appears in spatial reconstruction, where Dice improves from 0.84 for FNO to 0.91 for PAT. One possible reason is structural: vertical stability depends on long range coupling between plasma and passive conductors, and PAT exposes electromagnetic matrix embeddings directly to the attention block. This explanation remains an interpretation until component ablations isolate the individual contributions.

\label{sec:data_efficiency}

Moreover, we point out that high fidelity stability calculations cost $\sim$3~min per equilibrium, so data efficiency has practical value. We trained all architectures on progressively smaller fractions of the full training set (10\%, 25\%, 50\%, 75\%, 100\%) and evaluated on the fixed test set.

\begin{figure}[htbp]
\centering
\includegraphics[width=0.48\textwidth]{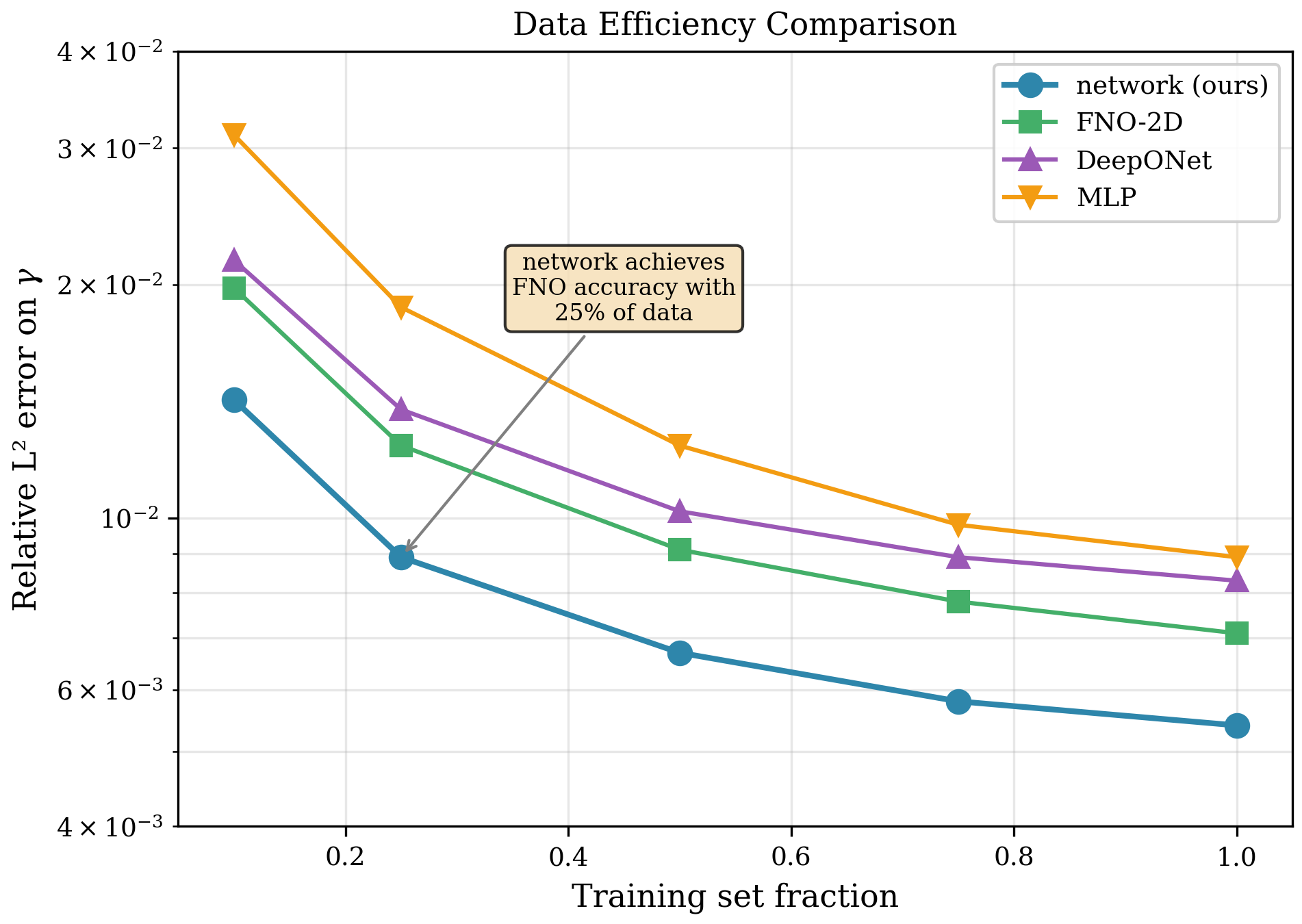}
\caption{Data efficiency comparison. PAT reaches approximately the full data FNO error using 25\% of the training set, while the scalar MLP saturates at higher error.}
\label{fig:data_efficiency}
\end{figure}

PAT also exhibits favorable data efficiency: at 25\% of training data (${\sim}3{,}500$ equilibria), it achieves $L^2(\hat{\gamma}) = 0.0073$, comparable to FNO's full data performance ($L^2 = 0.0071$). The learning curves in Figure~\ref{fig:data_efficiency} show that the MLP baseline saturates at $L^2 \approx 0.009$ regardless of data volume, consistent with a scalar input bottleneck. Under a fixed per equilibrium labeling cost assumption, the PAT data reduction would correspond to roughly 525 fewer CPU hours of MEQ-FGE-L labeling.

% Parameter definitions listed in Appendix A.

\section{Uncertainty Quantification and Out of Distribution Detection}
\label{sec:uncertainty}

Uncertainty estimates are useful only if they identify cases where the surrogate is likely to fail. We therefore test whether ensemble spread can flag shifted equilibria before they are trusted by a controller.

\label{sec:ensemble_method}

A 5-member deep ensemble\cite{lakshminarayanan2017simple} is trained with identical architecture but different random seeds. The ensemble mean serves as the point prediction; the ensemble standard deviation provides a proxy for epistemic uncertainty, $\sigma_{\text{epistemic}}$. Intervals are heuristic epistemic intervals around the ensemble mean, using Student's $t$-distribution ($N_e{-}1 = 4$ degrees of freedom) to account for the small ensemble size. They do not include EFIT reconstruction uncertainty, MEQ-FGE-L model form uncertainty, vessel model uncertainty, or synthetic SPARC uncertainty.

Calibration is assessed via a regression expected calibration error (ECE)\cite{guo2017calibration}: nominal central prediction intervals are binned by confidence level, and ECE is the mean absolute difference between nominal coverage and empirical coverage over those bins. No post hoc recalibration is applied. Table~\ref{tab:calibration} gives coverage and ECE for C-Mod, SPARC, and the SPARC extrapolation region.

\begin{table}[htbp]
\centering
\caption{Uncertainty calibration metrics. Coverage indicates the fraction of MEQ-FGE-L labels within the specified prediction interval. *Extrapolation region: equilibria with $\kappa > 2.2$ or $\beta_p > 1.5$.}
\label{tab:calibration}
\begin{tabular}{lccc}
\toprule
Dataset & ECE & 90\% Coverage & 95\% Coverage \\
\midrule
C-Mod test & 0.031 & 89.2\% & 94.1\% \\
SPARC SSL test & 0.047 & 86.8\% & 92.3\% \\
SPARC extrapolation* & 0.089 & 81.4\% & 88.7\% \\
\bottomrule
\end{tabular}
\end{table}

Calibration degrades progressively from C-Mod to SPARC and then to the SPARC extrapolation region, with ECE increasing from 0.031 to 0.089. The trend is useful but not sufficient: intervals widen under shift, yet they still under cover the most extrapolative equilibria.

\label{sec:ood_detection}

To connect ensemble spread with distributional shift, we define an OOD score $s_{\text{OOD}} = \sigma_{\text{epistemic}} / \bar{\sigma}^{\text{val}}_{\text{epistemic}}$, normalizing by the validation set mean uncertainty. We then compute the Mahalanobis distance $D_M$ from the training centroid in the space of scalar input parameters.

\begin{figure}[htbp]
\centering
\includegraphics[width=0.45\textwidth]{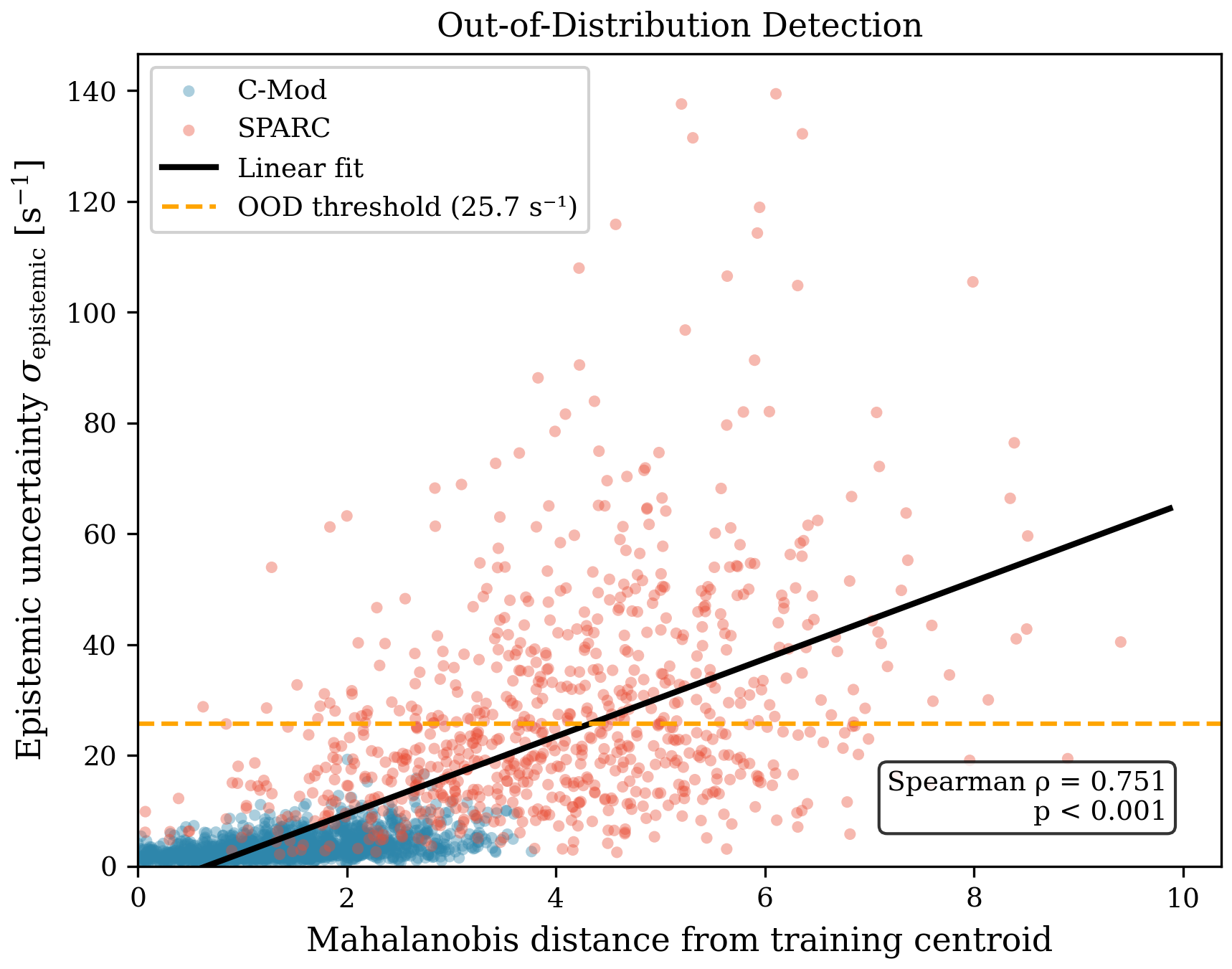}
\caption{Epistemic uncertainty vs.\ Mahalanobis distance from training centroid (blue: C-Mod; red: SPARC). Spearman $\rho = 0.73$. Dashed: $s_{\text{OOD}} = 2.0$.}
\label{fig:ood_detection}
\end{figure}

Epistemic uncertainty increases with Mahalanobis distance from the training centroid (Spearman $\rho = 0.73$, $p < 0.001$), as shown in Figure~\ref{fig:ood_detection}. SPARC extrapolation equilibria cluster at $D_M > 4$ and $\sigma_{\text{epistemic}} > 50$~s$^{-1}$. Input space distance can miss shifts that are visible only in mesh or latent features, so this score should be treated as a warning signal rather than a certificate.

The score therefore provides an empirical trigger for additional scrutiny. Equilibria flagged with $s_{\text{OOD}} > 2.5$ should trigger conservative control action, independent monitoring, or slower MEQ-FGE-L evaluation when latency permits. RZIP can serve as a low latency comparator, but it is not a higher fidelity fallback in the high elongation regimes where nonrigid effects matter.

\label{sec:uncertainty_error}

To test whether epistemic uncertainty is operationally useful, we compare $\sigma_{\text{epistemic}}$ with observed prediction error on the test set. Equilibria are grouped into quartiles by uncertainty, and MAE is computed within each bin.

Prediction error rises monotonically with ensemble spread in this test set. The lowest uncertainty quartile ($\sigma_{\text{epistemic}} < 5$~s$^{-1}$) has MAE 3.1~s$^{-1}$, while the highest uncertainty quartile ($\sigma_{\text{epistemic}} > 25$~s$^{-1}$) reaches 18.3~s$^{-1}$. The result motivates uncertainty gated review, but deployment would require selective risk curves or large residual detection AUROC/AUPRC to quantify how many dangerous cases are caught at a given deferral rate.

%%%%%%%%%%%%%%%%%%%%%%%%%%%%%%%%%%%%%%%%%%%%%%%%%%%%%%%%%%%%%
\section{Scalar Feature Attribution}
\label{sec:interpretability}

\label{sec:feature_importance}

\begin{figure}[htbp]
\centering
\includegraphics[width=0.48\textwidth]{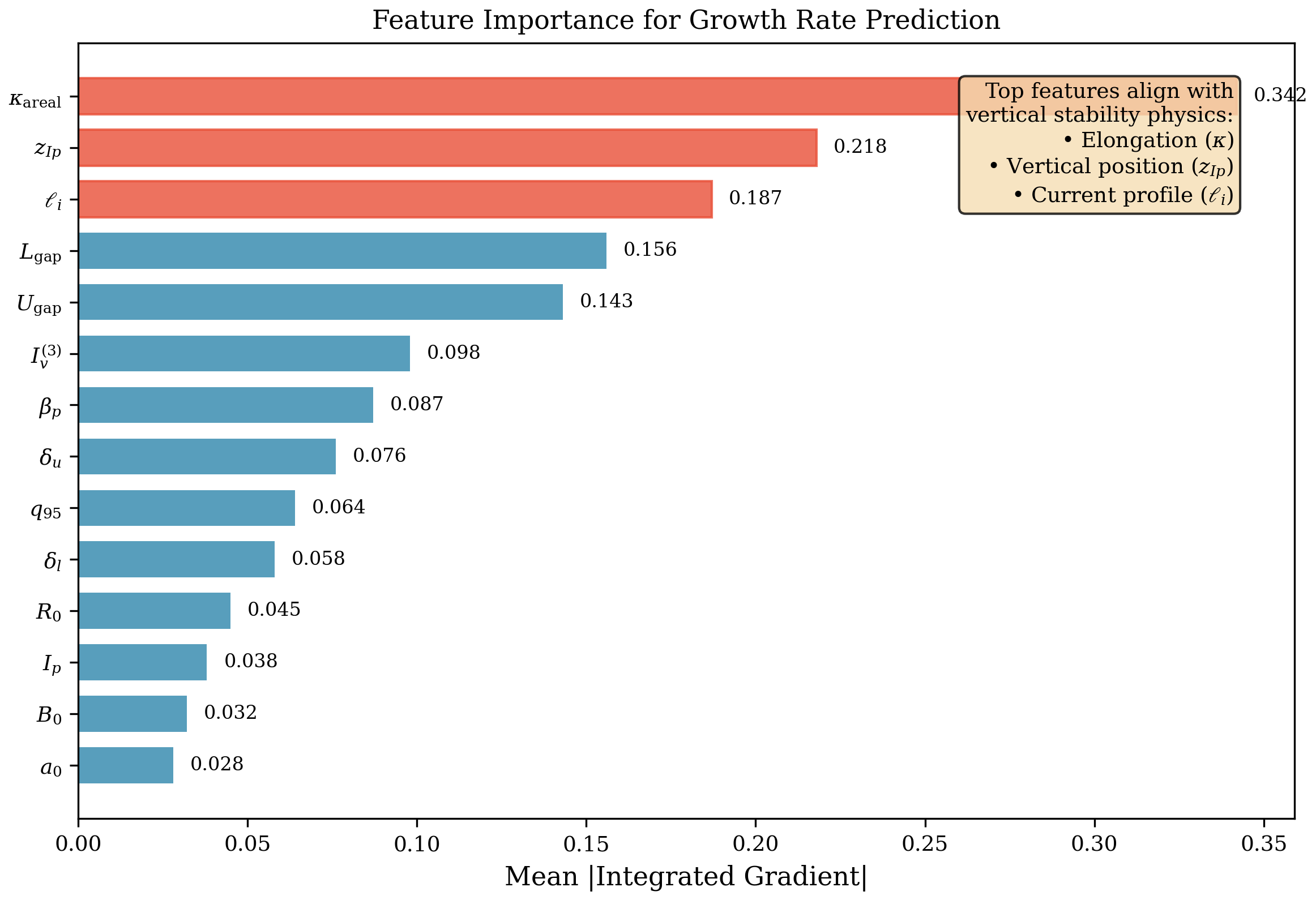}
\caption{Integrated gradient feature importance for growth rate prediction. Shape, vertical position, current profile, and wall gap features dominate the scalar sensitivity ranking.}
\label{fig:feature_importance}
\end{figure}

Integrated gradients (IG)\cite{sundararajan2017axiomatic} provide a scalar feature view of the same model, conditional on the trained surrogate and the input distribution. This ranking excludes mesh field channels and matrix token pathways, so it should not be read as a full attribution of every input route. Within that scalar subset, elongation dominates ($\kappa_{\text{areal}}$, mean $|\text{IG}| = 0.342$), followed by vertical position ($z_{\text{Ip}}$, 0.218) and internal inductance ($\ell_i$, 0.187). Plasma wall gaps rank fourth and fifth. Table~\ref{tab:feature_importance} lists the top ten scalar features with bootstrap confidence intervals, and Figure~\ref{fig:feature_importance} shows the same ranking visually. Here $I_v^{(3)}$ denotes the third passive vessel current channel in the ordered vessel current input vector. The ordering is stable across alternative baselines (magnitude changes $<8\%$) and agrees with the Sobol analysis above, making the sensitivities physically plausible without proving that they are causal.

\begin{table}[htbp]
\centering
\caption{Top ten input features by mean absolute integrated gradient (IG) across the test set. Bootstrap 95\% CIs computed with $B=1{,}000$ resamples.}
\label{tab:feature_importance}
\begin{tabular}{clcc}
\toprule
Rank & Feature & Mean $|\text{IG}|$ & 95\% CI \\
\midrule
1 & $\kappa_{\text{areal}}$ & 0.342 & [0.328, 0.356] \\
2 & $z_{\text{Ip}}$ & 0.218 & [0.204, 0.232] \\
3 & $\ell_i$ & 0.187 & [0.175, 0.199] \\
4 & $L_{\text{gap}}$ & 0.156 & [0.144, 0.168] \\
5 & $U_{\text{gap}}$ & 0.143 & [0.131, 0.155] \\
6 & $I_v^{(3)}$ & 0.098 & [0.088, 0.108] \\
7 & $\beta_p$ & 0.087 & [0.078, 0.096] \\
8 & $\delta_u$ & 0.076 & [0.067, 0.085] \\
9 & $q_{95}$ & 0.064 & [0.056, 0.072] \\
10 & $\delta_\ell$ & 0.058 & [0.050, 0.066] \\
\bottomrule
\end{tabular}
\end{table}

\section{Cross Device Generalization: Transfer from C-Mod to SPARC}
\label{sec:cross_device}

Cross device generalization is assessed through dimensionless normalization and zero shot transfer experiments. C-Mod and SPARC differ considerably in engineering parameters, wall geometry, coil layout, and toroidal field, while some dimensionless plasma parameters relevant to vertical stability partially overlap. Internal inductance distributions nearly coincide ($D_{\text{KL}} = 0.08$~nats), whereas elongation shows a moderate shift ($D_{\text{KL}} = 0.42$~nats). That overlap provides a limited basis for transfer, but wall and passive structure differences remain central to the problem.

\label{sec:dimensionless}

Predicting the dimensionless product $\gamma^* = \gamma\tau_w$ instead of $\gamma$ normalizes by a device specific wall time scale ($\tau_w^{\text{C-Mod}} \approx 15$~ms vs.\ $\tau_w^{\text{SPARC}} \approx 8$~ms). It does not remove the full multi conductor wall spectrum, coil coupling, passive structure geometry, or possible 3D vessel asymmetries. Even so, this normalization improves SPARC $R^2$ from 0.940 to 0.964 and reduces MAE from 12.7 to 8.9~s$^{-1}$ in the present two device split, making the dimensionless formulation preferable for this setting.

\label{sec:zero_shot_transfer}

To test generalization without SPARC labels, we trained PAT exclusively on C-Mod data and evaluated it on a SPARC benchmark overlap subset. The subset contains SPARC equilibria whose scalar parameters overlap the  C-Mod supported range closely enough for a controlled comparison. This is distinct from the full SPARC SSL test set, where the jointly trained model gives MAE 12.7~s$^{-1}$.

The benchmark overlap comparison in Table~\ref{tab:zero_shot_transfer} separates zero shot transfer from joint C-Mod/SPARC training.

\begin{table}[htbp]
\centering
\caption{Zero shot transfer vs.\ joint training on the SPARC benchmark overlap subset. The joint training MAE differs from Table~\ref{tab:bootstrap_ci} because Table~\ref{tab:bootstrap_ci} reports the full SPARC SSL test set.}
\label{tab:zero_shot_transfer}
\begin{tabular}{lcc}
\toprule
Metric & C-Mod Only & Joint Training \\
\midrule
SPARC MAE (s$^{-1}$) & 18.7 & 8.1 \\
SPARC $R^2$ & 0.78 & 0.94 \\
Mean $\sigma_{\text{epistemic}}$ (s$^{-1}$) & 24.3 & 7.6 \\
OOD flag rate ($s_{\text{OOD}} > 2$) & 67\% & 12\% \\
\bottomrule
\end{tabular}
\end{table}

The zero shot model achieves $R^2 = 0.78$, a substantial degradation from the jointly trained $R^2 = 0.94$, with MAE increasing by a factor of 2.3 on this subset. The degradation is accompanied by elevated uncertainty: mean $\sigma_{\text{epistemic}}$ is 3.2$\times$ higher, and 67\% of SPARC equilibria are flagged OOD. The high flag rate reflects distributional mismatch; the lower 12\% rate under joint training reflects that SPARC examples are then represented in the training distribution.

These results carry two implications. Synthetic SPARC data are needed to reach MEQ-FGE-L solver fidelity metrics in the SPARC like domain; extrapolation from C-Mod gives order of magnitude estimates but not the precision required near operational thresholds. At the same time, ensemble uncertainty identifies many zero shot predictions as unreliable, supporting its use as a warning signal rather than as a guarantee.

For SPARC commissioning studies, a tiered strategy can be tested: use PAT directly when $s_{\text{OOD}} < 1.5$, add enhanced monitoring for $1.5 \leq s_{\text{OOD}} < 2.5$, trigger conservative action or independent monitoring for $s_{\text{OOD}} \geq 2.5$, use RZIP only as a low latency comparator, and run MEQ-FGE-L when latency permits. The thresholds are empirical, not certified.

%%%%%%%%%%%%%%%%%%%%%%%%%%%%%%%%%%%%%%%%%%%%%%%%%%%%%%%%%%%%%
\section{Operational Limitations}
\label{sec:failure_analysis}

Average error is not the only quantity that matters for operation. The key question is whether high risk false negatives can be detected before they enter the control loop. We therefore examine above threshold fast growth detection and stratify errors by regime and magnetic configuration.

\label{sec:binary_classification}

The surrogate can also be used as an above threshold fast growth alarm. The task is not to decide whether the plasma is linearly stable, since any positive growth rate is unstable in the linearized model, but to determine whether $\gamma$ exceeds a threshold that would demand enhanced vertical stability response. The costs are asymmetric: a false negative misses a fast growth case, while a false positive triggers conservative control action. Recall oriented metrics are therefore emphasized.

We evaluate at the illustrative operating threshold $\gamma_{\text{crit}} = 100$~s$^{-1}$, separating routine margin cases from cases that would likely require stronger vertical stability response. The resulting classification metrics are given in Table~\ref{tab:binary_classification}.

\begin{table}[htbp]
\centering
\caption{Binary classification metrics at $\gamma_{\text{crit}} = 100$~s$^{-1}$. F$_2$-score weights recall twice as heavily as precision, reflecting the higher cost of false negatives in safety critical operation.}
\label{tab:binary_classification}
\begin{tabular}{lcc}
\toprule
Metric & C-Mod & SPARC \\
\midrule
AUC ROC & 0.994 & 0.981 \\
Sensitivity (TPR) & 97.8\% & 94.2\% \\
Specificity (TNR) & 96.4\% & 91.8\% \\
False Negative Rate & 2.2\% & 5.8\% \\
F$_2$-Score & 0.972 & 0.943 \\
\bottomrule
\end{tabular}
\end{table}

The C-Mod false negative rate of 2.2\% means approximately 1 in 45 above threshold equilibria would be missed at this threshold. Most false negatives also have elevated uncertainty ($s_{\text{OOD}} > 1.8$), so uncertainty gating can catch many, but not all, of the dangerous misses. A smaller subset remains confidently wrong; in SPARC, 11\% of false negatives have $s_{\text{OOD}} < 1.8$, which is why PAT cannot serve as a sole safety barrier.

The higher SPARC FNR (5.8\%) reflects the extrapolation regime. In threshold sweeps, lowering $\gamma_{\text{crit}}$ to 80~s$^{-1}$ reduces missed above threshold fast growth cases at the cost of more false positives. That tradeoff should be tuned with the control policy and disruption mitigation trigger, not chosen from surrogate metrics alone.

\label{sec:error_stratification}

For the subset with complete regime and magnetic configuration labels, Table~\ref{tab:error_stratification} stratifies MAE by operating regime and topology. The subset is smaller than the full growth rate test set because not every synthetic scenario has a one to one confinement regime annotation.

\begin{table}[htbp]
\centering
\caption{MAE stratified by confinement regime and magnetic configuration for the labeled regime subset. Training set representation ($N_{\text{train}}$) contextualizes accuracy differences.}
\label{tab:error_stratification}
\begin{tabular}{lccc}
\toprule
Category & $N_{\text{test}}$ & $N_{\text{train}}$ & MAE (s$^{-1}$) \\
\midrule
Ohmic & 412 & 2{,}748 & 4.2 \\
L-mode & 687 & 4{,}580 & 5.1 \\
I-mode & 234 & 1{,}560 & 6.3 \\
H-mode & 521 & 3{,}473 & 6.8 \\
\midrule
Lower Single Null (LSN) & 1{,}124 & 7{,}493 & 5.2 \\
Upper Single Null (USN) & 489 & 3{,}260 & 5.7 \\
Double Null (DN) & 241 & 1{,}608 & 7.4 \\
\bottomrule
\end{tabular}
\end{table}

\begin{figure*}[htbp]
\centering
\includegraphics[width=\textwidth]{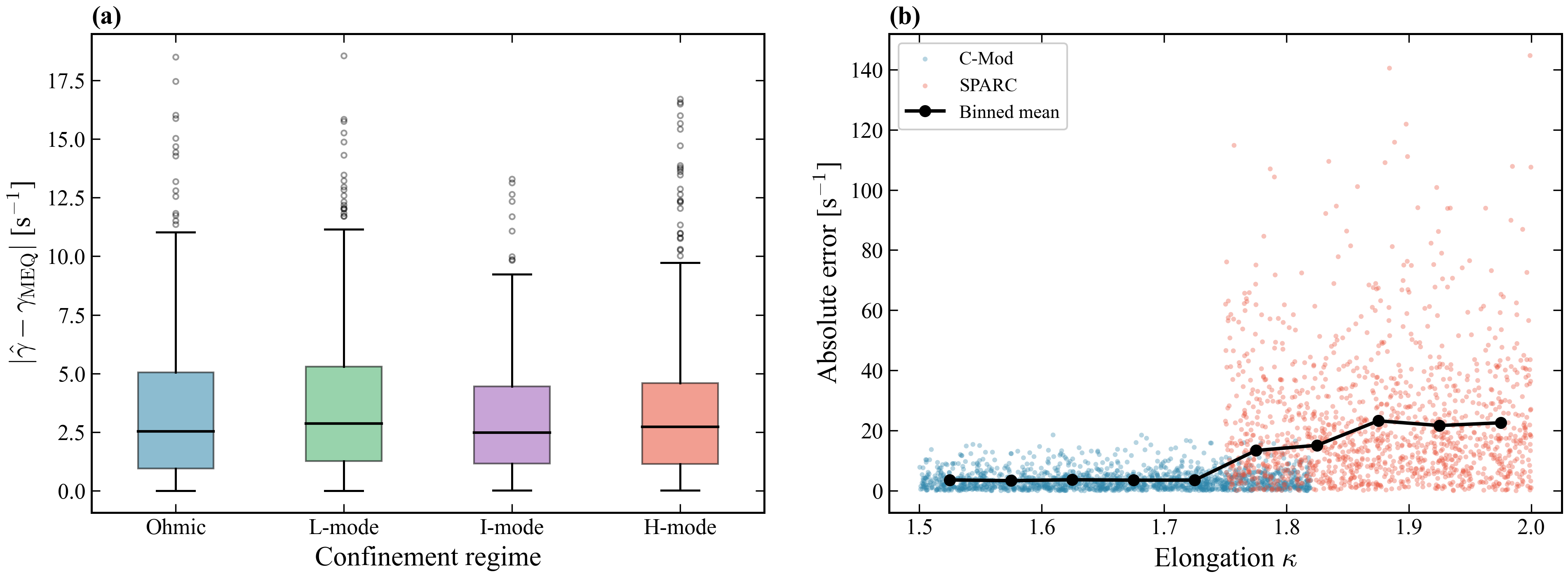}
\caption{(a) Absolute error by confinement regime. (b) Error vs.\ $\kappa$ (blue: C-Mod; red: SPARC; black: binned mean). Error rises above $\kappa \approx 1.85$.}
\label{fig:error_by_regime}
\end{figure*}

The stratification in Figure~\ref{fig:error_by_regime}(a) shows a consistent trend toward larger errors in higher performance regimes: MAE rises from 4.2~s$^{-1}$ for Ohmic cases to 6.8~s$^{-1}$ for H-mode cases. Double null configurations have the largest MAE, 7.4~s$^{-1}$. Geometry, profile gradients, reconstruction uncertainty, shape covariance, and data imbalance are all plausible contributors; DN cases contain two X points and make up only 13\% of the training set.

Elongation is the clearest error amplifier in Figure~\ref{fig:error_by_regime}(b). C-Mod equilibria below $\kappa=1.82$ maintain low errors, while the plotted SPARC subset shows increasing scatter above $\kappa \approx 1.85$. Because the full SPARC library extends to $\kappa=2.35$, this trend argues for targeted high-$\kappa$ double null labeling rather than broad uniform augmentation.

\section{Implications for Real Time Deployment}
\label{sec:deployment_implications}

Taken together, the reported accuracy, uncertainty, and latency support PAT as a fast advisory layer for solver labeled equilibria. On one CPU core, the single model chained pipeline is compatible with 50 to 100~ms control cycles used in many present day settings. A five member ensemble requires either parallel execution or a longer budget, and meeting a 10~ms target requires parallel CPU execution or GPU deployment. The timing implication is therefore hardware specific: PAT removes the minutes long MEQ-FGE-L bottleneck, but the final control loop budget still depends on EFIT latency, data transport, ensemble size, and fallback policy.

The second implication concerns data acquisition. Errors concentrate in high-$\kappa$ SPARC and double null cases, not uniformly across the operating space. Additional MEQ-FGE-L labeling should therefore prioritize high elongation, small gap, double null scenarios rather than simply increasing the training set uniformly. This targeted strategy would test whether the current error pattern comes from geometric complexity, sparse representation, or both.

The third implication is procedural. Ensemble uncertainty and OOD scores can triage predictions, but they are empirical alarms rather than guarantees. Low uncertainty PAT predictions can support fast screening and control cycle monitoring. High uncertainty predictions should trigger conservative action, reduced reliance on the surrogate, or slower physics based evaluation when time permits. For safety critical use, PAT should sit inside a defense in depth architecture with an independent monitor.

%%%%%%%%%%%%%%%%%%%%%%%%%%%%%%%%%%%%%%%%%%%%%%%%%%%%%%%%%%%%
\section{Conclusion}
\label{sec:conclusion}

We have described a physics attention surrogate that reproduces MEQ-FGE-L vertical growth rates with errors of a few s$^{-1}$ on C-Mod and ${\sim}13$~s$^{-1}$ on the full SPARC SSL test set. Normalized MEQ-FGE-L spatial eigenfunctions are reproduced at the ${\sim}5\%$ level. Single core inference is 35~ms, while parallel CPU or GPU deployment reaches the 10~ms scale.

One natural future use is a proximity monitor for loss of vertical control. A real time estimate of $\gamma$, together with uncertainty and OOD score, could be converted into a headroom measure that asks how far the present equilibrium is from a growth rate or $\gamma\tau_w$ limit where feedback margin becomes small. That signal could inform active shape feedback by penalizing requested changes in elongation, triangularity, or plasma wall gap that increase the predicted growth rate, or by selecting nearby shape targets that preserve vertical stability margin. In this role PAT would provide a fast growth rate aware constraint for shape control and scenario planning, while the actuator command would remain with validated plasma control algorithms and independent safety logic.

The practical role for PAT is therefore bounded but useful: fast screening, advisory monitoring, and uncertainty gated fallback inside a larger control architecture. Before deployment, MEQ-FGE-L should be benchmarked against experimental growth rate estimates and nonlinear MHD simulations, and PAT should be paired with an independent safety monitor. Planned extensions include temporal modeling for predictive control\cite{cotsim1} and multi device generalization to DIII-D and ITER.

\begin{acknowledgments}
This work was funded by Commonwealth Fusion Systems (CFS). The first author would like to thank Pinak Mandal for suggesting physics attention architecture `` Transolvers''. Simulations which were performed on the MIT-PSFC partition of the Engaging cluster at the MGHPCC facility (www.mghpcc.org). Alcator C-Mod data used in this study were generated with support from DOE Award DE-FC02-99ER54512. Initial testing of model simulations were done on MIT-PSFC GPU cluster: NVIDIA Tesla V100s with 32 GB HBM2. 
\end{acknowledgments}

\section*{Data Availability Statement}

The data that support the findings of this study are available
upon reasonable request from the authors.

\appendix
\section{MLP Baseline Model} \label{appendixA}

The MLP baseline\cite{bishop1995neural} predicts $\hat{\gamma}^{\text{MLP}}$ from 17 scalar equilibrium parameters: shape descriptors ($\kappa$, $\delta_u$, $\delta_\ell$), current profile ($I_p$, $\ell_i$, $q_{95}$), pressure ($\beta_p$, $W_{\text{diag}}$), geometry ($R_0$, $a_0$, gaps), and coil/vessel currents. It does not use mesh resolved data or electromagnetic coupling tensors.

The architecture consists of four hidden layers $[256, 256, 128, 64]$ with ReLU activations, selected via random hyperparameter search with Weights \& Biases\cite{wandb}. It was trained on identical splits and hardware (A100 GPU, $<1$~h, FP32) with the same normalization as PAT.

The MLP achieves $R^2 = 0.85$ on the combined test set. The $L^2$ error saturates at ${\approx}0.009$ regardless of data volume, consistent with a fundamental information bottleneck in the scalar input representation.

\bibliography{aipsamp}

@PREAMBLE{
 "\providecommand{\noopsort}[1]{}" 
 # "\providecommand{\singleletter}[1]{#1}%" 
}

@article{Heiss2025FGE,
  title   = {FGE: A Fast Free-Boundary Grad-Shafranov Evolutive Solver},
  author  = {Heiß, Cosmas and Merle, Antoine and Carpanese, Francesco and Felici, Federico and Donner, Craig and Marchioni, Stefano and Mari, Alessandro and Sauter, Olivier},
  journal = {arXiv preprint arXiv:2512.06847},
  year    = {2025},
  note    = {Part of the MEQ (Matlab Equilibrium) suite}
}

@inproceedings{sundararajan2017axiomatic,
  author    = {Sundararajan, Mukund and Taly, Ankur and Yan, Qiqi},
  title     = {{Axiomatic Attribution for Deep Networks}},
  booktitle = {Proceedings of the 34th International Conference on Machine Learning},
  series    = {Proceedings of Machine Learning Research},
  volume    = {70},
  pages     = {3319--3328},
  year      = {2017},
  publisher = {PMLR},
  url       = {https://proceedings.mlr.press/v70/sundararajan17a.html}
}

@inproceedings{paszke2019pytorch,
  author    = {Paszke, Adam and Gross, Sam and Massa, Francisco and Lerer, Adam
               and Bradbury, James and Chanan, Gregory and Killeen, Trevor
               and Lin, Zeming and Gimelshein, Natalia and Antiga, Luca
               and Desmaison, Alban and Kopf, Andreas and Yang, Edward
               and DeVito, Zachary and Raison, Martin and Tejani, Alykhan
               and Chilamkurthy, Sasank and Steiner, Benoit and Fang, Lu
               and Bai, Junjie and Chintala, Soumith},
  title     = {{PyTorch: An Imperative Style, High-Performance Deep Learning Library}},
  booktitle = {Advances in Neural Information Processing Systems},
  volume    = {32},
  year      = {2019},
}

@article{lakshminarayanan2017simple,
  author    = {Lakshminarayanan, Balaji and Pritzel, Alexander and Blundell, Charles},
  title     = {{Simple and Scalable Predictive Uncertainty Estimation using Deep Ensembles}},
  journal   = {arXiv preprint},
  eprint    = {1612.01474},
  year      = {2017},
  note      = {arXiv:1612.01474 [stat.ML]},
  url       = {https://arxiv.org/abs/1612.01474}
}

@inproceedings{loshchilov2016sgdr,
  author    = {Loshchilov, Ilya and Hutter, Frank},
  title     = {{SGDR: Stochastic Gradient Descent with Warm Restarts}},
  booktitle = {International Conference on Learning Representations},
  year      = {2017},
  note      = {arXiv:1608.03983},
  url       = {https://arxiv.org/abs/1608.03983}
}

@inproceedings{guo2017calibration,
  author    = {Guo, Chuan and Pleiss, Geoff and Sun, Yu and Weinberger, Kilian Q.},
  title     = {{On Calibration of Modern Neural Networks}},
  booktitle = {Proceedings of the 34th International Conference on Machine Learning},
  pages     = {1321--1330},
  year      = {2017},
  editor    = {Precup, Doina and Teh, Yee Whye},
  volume    = {70},
  series    = {Proceedings of Machine Learning Research},
  publisher = {PMLR},
  url       = {https://proceedings.mlr.press/v70/guo17a.html}
}

@article{sobol2001global,
  author  = {Sobol’, Ilya M.},
  title   = {{Global sensitivity indices for nonlinear mathematical models and their Monte Carlo estimates}},
  journal = {Mathematics and Computers in Simulation},
  volume  = {55},
  number  = {1--3},
  pages   = {271--280},
  year    = {2001},
  doi     = {10.1016/S0378-4754(00)00270-6}
}

@article{lu2021learning,
  author  = {Lu, Lu and Jin, Pengzhan and Pang, Guofei and Zhang, Zhongqiang and Karniadakis, George Em},
  title   = {{Learning nonlinear operators via DeepONet based on the universal approximation theorem of operators}},
  journal = {Nature Machine Intelligence},
  volume  = {3},
  number  = {3},
  pages   = {218--229},
  year    = {2021},
  doi     = {10.1038/s42256-021-00302-5}
}

@phdthesis{Marchioni2024,
  author = {Stefano Marchioni},
  title = {Vertical Instability Studies in the TCV Tokamak and Development and Application of Multimachine Real-Time Proximity Control Strategies},
  school = {EPFL},
  year = {2024},
  note = {Thèse No. 10943}
}

@article{Li2020fno,
  author = {Zongyi Li and others},
  title = {Fourier Neural Operator for Parametric Partial Differential Equations},
  journal = {arXiv preprint arXiv:2010.08895},
  year = {2020}
}

@article{Gopakumar_2024,
doi = {10.1088/1741-4326/ad313a},
url = {https://dx.doi.org/10.1088/1741-4326/ad313a},
year = {2024},
month = {apr},
publisher = {IOP Publishing},
volume = {64},
number = {5},
pages = {056025},
author = {Gopakumar, Vignesh and Pamela, Stanislas and Zanisi, Lorenzo and Li, Zongyi and Gray, Ander and Brennand, Daniel and Bhatia, Nitesh and Stathopoulos, Gregory and Kusner, Matt and Peter Deisenroth, Marc and Anandkumar, Anima and the JOREK Team and MAST Team},
title = {Plasma surrogate modelling using Fourier neural operators},
journal = {Nuclear Fusion}
}

@article{decaf,
    author = {Sabbagh, S. A. and Berkery, J. W. and Park, Y. S. and Butt, J. and Riquezes, J. D. and Bak, J. G. and Bell, R. E. and Delgado-Aparicio, L. and Gerhardt, S. P. and Ham, C. J. and Hollocombe, J. and Lee, J. W. and Kim, J. and Kirk, A. and Ko, J. and Ko, W. H. and Kogan, L. and LeBlanc, B. P. and Lee, J. H. and Thornton, A. and Yoon, S. W.},
    title = {Disruption event characterization and forecasting in tokamaks},
    journal = {Physics of Plasmas},
    volume = {30},
    number = {3},
    pages = {032506},
    year = {2023},
    month = {03},
    issn = {1070-664X},
    doi = {10.1063/5.0133825},
    url = {https://doi.org/10.1063/5.0133825},
    eprint = {https://pubs.aip.org/aip/pop/article-pdf/doi/10.1063/5.0133825/19824942/032506\_1\_online.pdf},
}

@INPROCEEDINGS{cotsim1,
  author={Pajares, Andres and Schuster, Eugenio},
  booktitle={2019 IEEE 58th Conference on Decision and Control (CDC)}, 
  title={Integrated Robust Control of Individual Scalar Variables in Tokamaks}, 
  year={2019},
  volume={},
  number={},
  pages={3233-3238},
  doi={10.1109/CDC40024.2019.9029195}}

@inproceedings{adam_optimizer,
  author = {Kingma, Diederik P. and Ba, Jimmy},
  booktitle = {ICLR },
  ee = {http://arxiv.org/abs/1412.6980},
  title = {Adam: A Method for Stochastic Optimization.},
  url = {http://dblp.uni-trier.de/db/conf/iclr/iclr2015.html#KingmaB14},
  year = 2015
}

@misc{wandb,
title = {Experiment Tracking with Weights and Biases},
year = {2020},
note = {Software available from wandb.com},
url={https://www.wandb.com/},
author = {Biewald, Lukas},
}

@Article{app10196683,
AUTHOR = {Murari, Andrea and Peluso, Emmanuele and Lungaroni, Michele and Rossi, Riccardo and Gelfusa, Michela and JET Contributors},
TITLE = {Investigating the Physics of Tokamak Global Stability with Interpretable Machine Learning Tools},
JOURNAL = {Applied Sciences},
VOLUME = {10},
YEAR = {2020},
NUMBER = {19},
ARTICLE-NUMBER = {6683},
URL = {https://www.mdpi.com/2076-3417/10/19/6683},
ISSN = {2076-3417},
DOI = {10.3390/app10196683}
}

@PhdThesis{francesco2021,
  title = {Development of Free-Boundary Equilibrium and Transport Solvers for Simulation and Real-Time Interpretation of Tokamak Experiments},
  author = {Francesco Carpanese},
  school = {EPFL},
  year = {2021},
  doi = {10.5075/epfl-thesis-7914},
  type = {PhD Thesis}
}

@inproceedings{morabito1997fuzzy,
  title={A fuzzy-neural approach to real time plasma boundary reconstruction in tokamak reactors},
  author={Morabito, Francesco Carlo and Versaci, Mario},
  booktitle={Proceedings of International Conference on Neural Networks (ICNN'97)},
  volume={1},
  pages={43--47},
  year={1997},
  organization={IEEE}
}

@article{Montes_2019,
doi = {10.1088/1741-4326/ab1df4},
url = {https://dx.doi.org/10.1088/1741-4326/ab1df4},
year = {2019},
month = {jul},
publisher = {IOP Publishing},
volume = {59},
number = {9},
pages = {096015},
author = {Montes, K.J. and Rea, C. and Granetz, R.S. and Tinguely, R.A. and Eidietis, N. and Meneghini, O.M. and Chen, D.L. and Shen, B. and Xiao, B.J. and Erickson, K. and Boyer, M.D.},
title = {Machine learning for disruption warnings on Alcator C-Mod, DIII-D, and EAST},
journal = {Nuclear Fusion}
}

@article{Li2021geo,
  author = {Zongyi Li and others},
  title = {Fourier Neural Operator with Learned Deformations for PDEs on General Geometries},
  journal = {arXiv preprint arXiv:2207.05209},
  year = {2021}
}

@book{bishop1995neural,
  title={Neural networks for pattern recognition},
  author={Bishop, Christopher M},
  year={1995},
  publisher={Oxford university press}
}

@inproceedings{Wu2023latent,
  author = {Haixu Wu and others},
  title = {Solving High-Dimensional PDEs with Latent Spectral Models},
  booktitle = {ICML},
  year = {2023}
}

@inproceedings{Wu2024,
  author = {Haixu Wu and Huakun Luo and Haowen Wang and Jianmin Wang and Mingsheng Long},
  title = {Transolver: A Fast Transformer Solver for PDEs on General Geometries},
  booktitle = {ICML},
  year = {2024}
}

@article{Grad1967,
  author={Grad, H.},
  title={Toroidal containment of a plasma},
  journal={Physics of Fluids},
  volume={10},
  number={1},
  pages={137--154},
  year={1967},
  doi={10.1063/1.1761927}
}

@unpublished{wai2025,
  author = "Wai, J. et al.",
  title  = "Evolution: a revised theory",
  year   = 2006,
  note = "in-preparation"
}

@article{Hofmann_1997,
doi = {10.1088/0029-5515/37/5/I10},
url = {https://dx.doi.org/10.1088/0029-5515/37/5/I10},
year = {1997},
month = {may},
publisher = {},
volume = {37},
number = {5},
pages = {681},
author = {F. Hofmann and M.J. Dutch and D.J. Ward and M. Anton and I. Furno and J.B. Lister and J.-M. Moret},
title = {Vertical instability in TCV: comparison of experimental and theoretical growth rates},
journal = {Nuclear Fusion}
}

@article{Humphreys_2009,
doi = {10.1088/0029-5515/49/11/115003},
url = {https://dx.doi.org/10.1088/0029-5515/49/11/115003},
year = {2009},
month = {sep},
publisher = {},
volume = {49},
number = {11},
pages = {115003},
author = {Humphreys, D.A. and Casper, T.A. and Eidietis, N. and Ferrara, M. and Gates, D.A. and Hutchinson, I.H. and Jackson, G.L. and Kolemen, E. and Leuer, J.A. and Lister, J. and LoDestro, L.L. and Meyer, W.H. and Pearlstein, L.D. and Portone, A. and Sartori, F. and Walker, M.L. and Welander, A.S. and Wolfe, S.M.},
title = {Experimental vertical stability studies for ITER performance and design guidance},
journal = {Nuclear Fusion}
}

@article{Olofsson_2022,
doi = {10.1088/1361-6587/ac6ffd},
url = {https://dx.doi.org/10.1088/1361-6587/ac6ffd},
year = {2022},
month = {may},
publisher = {IOP Publishing},
volume = {64},
number = {7},
pages = {072001},
author = {Olofsson, K E J},
title = {Fast calculation of the tokamak vertical instability},
journal = {Plasma Physics and Controlled Fusion}
}

@article{welander_nonrigid_2005,
	title = {Nonrigid, {Linear} {Plasma} {Response} {Model} {Based} on {Perturbed} {Equilibria} for {Axisymmetric} {Tokamak} {Control} {Design}},
	volume = {47},
	issn = {1536-1055},
	url = {https://doi.org/10.13182/FST05-A778},
	doi = {10.13182/FST05-A778},
	number = {3},
	urldate = {2022-05-08},
	journal = {Fusion Science and Technology},
	author = {Welander, A. S. and Deranian, R. D. and Humphreys, D. A. and Leuer, J. A. and Walker, M. L.},
	month = apr,
	year = {2005},
	note = {Publisher: Taylor \& Francis
\_eprint: https://doi.org/10.13182/FST05-A778},
	pages = {763--767},
}

@article{humphreys_axisymmetric_1993,
	title = {Axisymmetric {Magnetic} {Control} {Design} in {Tokamaks} {Using} {Perturbed} {Equilibrium} {Plasma} {Response} {Modeling}},
	volume = {23},
	issn = {0748-1896},
	url = {https://doi.org/10.13182/FST93-A30146},
	doi = {10.13182/FST93-A30146},
	number = {2},
	urldate = {2020-06-09},
	journal = {Fusion Technology},
	author = {Humphreys, David A. and Hutchinson, Ian H.},
	month = mar,
	year = {1993},
	note = {Publisher: Taylor \& Francis
\_eprint: https://doi.org/10.13182/FST93-A30146},
	pages = {167--184},
}

@article{lazarus_control_1990,
	title = {Control of the vertical instability in tokamaks},
	volume = {30},
	url = {http://stacks.iop.org/0029-5515/30/i=1/a=010?key=crossref.1e0c01973636bdfad61f224a5e7a8015},
	doi = {10.1088/0029-5515/30/1/010},
	number = {1},
	journal = {Nuclear Fusion},
	author = {Lazarus, E A and Lister, J B and Neilson, G H},
	month = jan,
	year = {1990},
	pages = {111--141},
}

@article{leuer_passive_1989,
	title = {Passive {Vertical} {Stability} in the {Next} {Generation} {Tokamaks}},
	volume = {15},
	issn = {0748-1896},
	url = {https://doi.org/10.13182/FST89-A39747},
	doi = {10.13182/FST89-A39747},
	number = {2P2A},
	urldate = {2022-05-08},
	journal = {Fusion Technology},
	author = {Leuer, J. A.},
	month = mar,
	year = {1989},
	note = {Publisher: Taylor \& Francis
\_eprint: https://doi.org/10.13182/FST89-A39747},
	pages = {489--494},
}

@article{rodriguez-fernandez_overview_2022,
	title = {Overview of the {SPARC} physics basis towards the exploration of burning-plasma regimes in high-field, compact tokamaks},
	volume = {62},
	issn = {0029-5515, 1741-4326},
	url = {https://iopscience.iop.org/article/10.1088/1741-4326/ac1654},
	doi = {10.1088/1741-4326/ac1654},
	number = {4},
	urldate = {2023-05-15},
	journal = {Nuclear Fusion},
	author = {Rodriguez-Fernandez, P. and Creely, A.J. and Greenwald, M.J. and Brunner, D. and Ballinger, S.B. and Chrobak, C.P. and Garnier, D.T. and Granetz, R. and Hartwig, Z.S. and Howard, N.T. and Hughes, J.W. and Irby, J.H. and Izzo, V.A. and Kuang, A.Q. and Lin, Y. and Marmar, E.S. and Mumgaard, R.T. and Rea, C. and Reinke, M.L. and Riccardo, V. and Rice, J.E. and Scott, S.D. and Sorbom, B.N. and Stillerman, J.A. and Sweeney, R. and Tinguely, R.A. and Whyte, D.G. and Wright, J.C. and Yuryev, D.V.},
	month = sep,
	year = {2022},
	pages = {042003},
}

@article{Nelson_2024,
doi = {10.1088/1741-4326/ad58f6},
url = {https://dx.doi.org/10.1088/1741-4326/ad58f6},
year = {2024},
month = {jun},
publisher = {IOP Publishing},
volume = {64},
number = {8},
pages = {086040},
author = {A.O. Nelson and D.T. Garnier and D.J. Battaglia and C. Paz-Soldan and I. Stewart and M. Reinke and A.J. Creely and J. Wai},
title = {Implications of vertical stability control on the SPARC tokamak},
journal = {Nuclear Fusion}
}

@article{Okabayashi_1974,
doi = {10.1088/0029-5515/14/2/011},
url = {https://dx.doi.org/10.1088/0029-5515/14/2/011},
year = {1974},
month = {apr},
publisher = {},
volume = {14},
number = {2},
pages = {263},
author = {M. Okabayashi and G. Sheffield},
title = {Vertical stability of elongated tokamaks},
journal = {Nuclear Fusion}
}

@article{Creely_2020, 
title={Overview of the SPARC tokamak}, volume={86}, DOI={10.1017/S0022377820001257}, number={5}, journal={Journal of Plasma Physics}, author={Creely, A. J. and Greenwald, M. J. and Ballinger, S. B. and Brunner, D. and Canik, J. and Doody, J. and Fülöp, T. and Garnier, D. T. and Granetz, R. and Gray, T. K. and et al.}, year={2020}, pages={865860502}}

@article{Marmar_2015,
doi = {10.1088/0029-5515/55/10/104020},
url = {https://dx.doi.org/10.1088/0029-5515/55/10/104020},
year = {2015},
month = {jul},
publisher = {IOP Publishing},
volume = {55},
number = {10},
pages = {104020},
author = {E.S. Marmar and S.G. Baek and H. Barnard and P. Bonoli and D. Brunner and J. Candy and J. Canik and R.M. Churchill and I. Cziegler and G. Dekow and L. Delgado-Aparicio and A. Diallo and E. Edlund and P. Ennever and I. Faust and C. Fiore and Chi Gao and T. Golfinopoulos and M. Greenwald and Z.S. Hartwig and C. Holland and A.E. Hubbard and J.W. Hughes and I.H. Hutchinson and J. Irby and B. LaBombard and Yijun Lin and B. Lipschultz and A. Loarte and R. Mumgaard and R.R. Parker and M. Porkolab and M.L. Reinke and J.E. Rice and S. Scott and S. Shiraiwa and P. Snyder and B. Sorbom and D. Terry and J.L. Terry and C. Theiler and R. Vieira and J.R. Walk and G.M. Wallace and A. White and D. Whyte and S.M. Wolfe and G.M. Wright and J. Wright and S.J. Wukitch and P. Xu},
title = {Alcator C-Mod: research in support of ITER and steps beyond},
journal = {Nuclear Fusion}
}

@article{Albanese_1989,
doi = {10.1088/0029-5515/29/6/011},
url = {https://dx.doi.org/10.1088/0029-5515/29/6/011},
year = {1989},
month = {jun},
publisher = {},
volume = {29},
number = {6},
pages = {1013},
author = {R. Albanese and E. Coccorese and G. Rubinacci},
title = {Plasma modelling for the control of vertical instabilities in tokamaks},
journal = {Nuclear Fusion}
}

@article{creely23,
    author = {Creely, A. J. and Brunner, D. and Mumgaard, R. T. and Reinke, M. L. and Segal, M. and Sorbom, B. N. and Greenwald, M. J.},
    title = {SPARC as a platform to advance tokamak science},
    journal = {Physics of Plasmas},
    volume = {30},
    number = {9},
    pages = {090601},
    year = {2023},
    month = {09},
    issn = {1070-664X},
    doi = {10.1063/5.0162457},
    url = {https://doi.org/10.1063/5.0162457},
    eprint = {https://pubs.aip.org/aip/pop/article-pdf/doi/10.1063/5.0162457/19982805/090601\_1\_5.0162457.pdf},
}

@article{Ward1992EffectsOP,
  title={Effects of plasma deformability on the feedback stabilization of axisymmetric modes in tokamak plasmas},
  author={David J. Ward and Stephen C. Jardin},
  journal={Nuclear Fusion},
  year={1992},
  volume={32},
  pages={973-994},
  url={https://api.semanticscholar.org/CorpusID:120496816}
}

@article{Ferrara2008PlasmaIA,
  title={Plasma inductance and stability metrics on Alcator C-Mod},
  author={M. Ferrara and Ian H. Hutchinson and Stephen M. Wolfe},
  journal={Nuclear Fusion},
  year={2008},
  volume={48},
  pages={065002},
  url={https://api.semanticscholar.org/CorpusID:123140894}
}

\end{document}